\documentclass[reprint,superscriptaddress,amsmath,amssymb,aps,preprintnumbers]{revtex4-1}
\usepackage{graphicx}
\usepackage{dcolumn}
\usepackage{bm}
\usepackage[usenames]{color}
\usepackage[normalem]{ulem}
\usepackage[colorlinks=true, pdfstartview=FitV, linkcolor=blue, citecolor=blue, urlcolor=blue]{hyperref}

\renewcommand{\sout}[1]{\bgroup \color{red} \ULdepth=-.5ex \ULset {#1}}
\newcommand{\comment}[1]{}
\graphicspath{{./Figs/}}
\newcommand{\average}[1]{\ensuremath{\langle#1\rangle}}

\newcommand{\tJo}{ \tilde{J}^{(1)} }
\newcommand{\tJt}{ \tilde{J}^{(2)} }
\newcommand{\ttau}{ \tilde{\tau} }
\newcommand{\teta}{ \tilde{\eta} }

\newcommand{\bx}{ \bm{x} }
\newcommand{\bk}{ \bm{k} }
\newcommand{\bxp}{ \bm{x}_\perp }

\newcommand{\bbpi}{ \bm{b}_{\perp,i} }
\newcommand{\refeq}[1]{(\ref{#1})}
\newcommand{\xm}{ x^- }
\newcommand{\xp}{ x^+ }

\begin{document}
\preprint{}
\title{
Simulation of 3+1D glasma in Milne coordinates I: Development of the framework}
\author{Hidefumi Matsuda}
\email{da.matsu.00.bbb.kobe@gmail.com}
\affiliation{Physics Department and Center for Particle Physics and Field Theory, Fudan University, Shanghai 200438, China}
\author{Xu-Guang Huang}
\email{huangxuguang@fudan.edu.cn}
\affiliation{Physics Department and Center for Particle Physics and Field Theory, Fudan University, Shanghai 200438, China}
\affiliation{Key Laboratory of Nuclear Physics and Ion-beam Application (MOE), Fudan University, Shanghai 200433, China}
\affiliation{Shanghai Research Center for Theoretical Nuclear Physics, National Natrual Science Foundation of China and Fudan University, Shanghai 200438, China}

\begin{abstract}
We propose a new numerical method for $3+1$D glasma simulation using Milne coordinates. 
We formulate the classical Yang-Mills field and $3$D classical color current on a lattice at the initial proper time, specified as a moment just before the collision of the two nuclei. By solving the evolution equations, we extract observables of the $3$D glasma at later times. We demonstrate the efficiency of our method in terms of numerical cost and apply it to the central collisions of Au-Au. We also discuss possible further improvements of our method.
\end{abstract}

\maketitle

\section{Introduction}
The experiment of relativistic heavy-ion collisions provides us the unique way to create deconfined quantum chromodynamics (QCD) matter of extraordinarily high temperature and density. Over the past few decades, many experimental results have indicated the emergence of a new state of matter, referred to as the quark-gluon plasma (QGP), in heavy-ion collisions, where quarks and gluons behave as a hydrodynamic fluid. The analysis of a relativistic heavy-ion collision requires different kinds of descriptions of the spacetime evolution of the system since the matter produced in the collision experiences varied stages in its evolution. The classical Yang-Mills (CYM) theory offers one of such descriptions. It can well describe the non-equilibrium evolution of the highly occupied gluonic system, called glasma, that appears immediately after the collision~\cite{MV1994,LMW1995,2Dglasma_KV,2Dglasma_KNR,2Dglasma_L,2Dglasma_LM,InstabilityCYM2006_RV,InstabilityCYM2008,InstabilityCYM2009,InstabilityCYM2012FG,InstabilityCYM2012BSSS,InstabilityCYM2013,PressureEG2013,Tsukiji2018,HM_2022}. The glasma simulation with the CYM field plays an important role in understanding the non-equilibrium stage between the moment of the collision and the onset of the hydrodynamic evolution of the QGP. In fact, the glasma simulation is widely used to establish the initial conditions for subsequent hydrodynamic evolution in the analysis of experimental data~\cite{IPglasma}. 

The theoretical background for why the CYM theory is a good description of the initial gluonic matter is based on the color glass condensate (CGC) picture, which stands as a valid description of the high-energy nucleus~\cite{MV1994,LMW1995}. In such a high-energy nucleus, the dominant degrees of freedom are soft gluons emitted from hard partons. The McLerran-Venugopalan (MV) model in the CGC effective theory describes the soft gluons as the CYM fields and the hard partons as their color sources. Consequently, the glasma generated in the collision of such high-energy nuclei can also be described well by the CYM field.

However, such a success of the glasma simulation has been largely limited by the boost invariance assumption, which shows good agreement with experimental data only around the midrapidity region. Recently, much attention has been paid to the $3+1$D glasma simulation beyond the boost invariance assumption that is necessary to understand observables across a broader region of rapidity~\cite{3DglasmaJIMWLK,3DglasmaJIMWLK2,3DglasmaCPIC,3DglasmaCPIC2,3DglasmaCPIC3,3DglasmaSS,3DglasmaAnalytic}. Different approaches for incorporating the rapidity dependence and their implementations have been considered. In Refs.~\cite{3DglasmaJIMWLK,3DglasmaJIMWLK2}, the authors consider the rapidity-dependent distribution of the classical color charges inside a single nucleus by numerically solving the JIMWLK equation~\cite{JIMWLK1997,JIMWLK1998,JIMWLK2001,JIMWLK2002,JIMWLK_1_2001,JIMWLK_2_2001,JIMWLK_LM},  while assuming that the color sources are static in time. In Refs~\cite{3DglasmaCPIC,3DglasmaCPIC2,3DglasmaCPIC3,3DglasmaSS}, the authors propose numerical simulation methods that focus on the recoil effect of the nuclei and track the dynamical evolution of the CYM field with the dynamical $3$D color current. The analytic analysis for the 3+1D glasma with the dynamical color current is also performed, employing the weak field approximation~\cite{3DglasmaAnalytic}.

The purpose of this study is to propose a new numerical simulation method for the $3+1$D glasma with the incorporation of the recoil effect, in which the classical color current is treated as a $3$D dynamical object. The initial conditions for the CYM field and classical color current are provided on a lattice before the collision occurs, and their discretized evolution equations are subsequently solved to determine their values at later times. Numerical simulations are performed in Milne coordinates, albeit with a difference from the usual Milne coordinates. Usually, the proper time $\tau=\sqrt{t^2-z^2}$ is introduced such that the collision of the two nuclei occurs at $\tau=0$. In contrast, we employ a modified Milne coordinates, $(\tilde{\tau}=\sqrt{\tilde{t}^2-z^2},x,y,\tilde{\eta}=(1/2)\ln[(\tilde{t}+z)/(\tilde{t}-z)])$, where $\tilde{t}$ is shifted from $t$ by a positive constant as $\tilde{t}=t+{\rm positive\ constant}$. Consequently, the two nuclei are still apart at the initial proper time $\tilde{\tau}_{\rm ini}$ which is taken as a sufficiently small number. We evaluate the physical quantities in the modified Milne coordinates and then transform the results to the usual Milne coordinates via a general coordinate transformation. The above strategy of giving initial conditions on a lattice, evolving them in time, and transforming the results into the usual Milne coordinates is analogous with that in Ref.~\cite{3DglasmaSS}, where the simulations are performed in Minkowski coordinates and the results are then transformed into the usual Milne coordinates. The advantage of using the Milne coordinates is the following. The numerical simulations on a finite lattice in Milne coordinates correspond to a longitudinally expanding system in terms of Minkowski coordinates due to the relation $z=\tilde{\tau} \sinh{\tilde{\eta}}$. As a result, numerical simulations in the Milne coordinates do not require a large lattice size in the $\tilde{\eta}$ direction, while numerical simulations in Minkowski coordinates require a system size in the $z$ direction large enough to include the outgoing nuclei within the lattice. Therefore, our new method is expected to cost lower numerical resources, which is important in actual applications since tracking the dynamical evolution of the $3$D glasma requires a lot of numerical resources. 

In Sec.~\ref{Sec:Method}, we present the formulation of the $3+1$D glasma on the lattice. 
In Sec.~\ref{Sec:result}, we present the numerical results. This section is divided into two parts. In the first part, we test the effectiveness of our numerical method. We check whether the continuity equations are violated under the evolution of the glasma and check the consistency with the method proposed in Ref.~\cite{3DglasmaSS} by comparing our results for the transverse pressure and the energy density in the local rest frame with theirs. In the second part, we simulate the dynamical evolution of the $3+1$D glasma using the setup that mimics the central  collisions of Au-Au at $\sqrt{s}=200$ GeV. We show the dynamical evolution of the energy density and address discussions about the obtained results. In Sec.~\ref{Summary}, we summarize our main results.

\section{Method}\label{Sec:Method}
We develop the numerical method for the $3+1$D glasma simulation in Milne coordinates in this section. This method is an extension of the description of the $2+1$D glasma using the MV model in the CGC effective theory. This section is organized as follows. In Sec.~\ref{Sec:2d}, we give a brief review of the description of the boost-invariant ($2+1$D) glasma using the MV model. In Sec.~\ref{Sec:3dcont}, we explain how to extend the $2+1$D glasma description to the $3+1$D glasma with the dynamical classical color current in continuous spacetime. In Sec.~\ref{Sec:3discr}, we show the formulation of the $3+1$D glasma on a discretized space and continuous proper time. In Sec.~\ref{Sec:EMtensor}, we define the energy-momentum tensor on a lattice.

\subsection{$2+1$D glasma in continuous spacetime}\label{Sec:2d}
Here we briefly review how the $2+1$D glasma is described using the MV model in the CGC effective theory. According to the CGC picture, the dominant degrees of freedom inside a relativistic nucleus are the soft gluons that are emitted from partons with large momenta. In the MV model, the soft and hard partons are separately treated in the classical approximation~\cite{MV1994,LMW1995}: The soft partons are described by the CYM field $A_\mu$ and the hard partons are described by the classical color current $J^\mu$, the source of the soft partons. Here, the classical color current $J^\mu$ moving toward the positive $z$ direction is given by the density of the sum of the color charges carried by hard partons located around $x$,
\begin{align}
J^\mu(x) = \frac{1}{g} \delta^{\mu,+} \rho(x^-,\bxp)\ ,\label{Eq:J_single}
\end{align}
where $x^{\mp}=(t \mp z)/\sqrt{2}$ are the light-cone coordinates and $\rho$ is the classical color charge density, randomly given according to a probability density $P[\rho]$ for each event. This color charge density $\rho$ is also assumed to be static, namely independent of $x^+$, which is reflected by the fact that the lifetime of the hard partons is much longer than that of the soft partons due to the time dilation. The soft CYM field emitted from the static sources is given by the solution to the classical equations of motion $[D_\mu, F^{\mu\nu}]=J^\nu$, and under the gauge condition $A_- = 0$, it has the following form~\cite{MV1994},
\begin{align}
A_\pm = 0\ ,\ \ A_i = \frac{i}{g} V \partial_i V^\dagger\ ,\label{Eq:A_single}
\end{align}
where $A_i$ is the transverse gauge field and $V^\dagger$ is the Wilson line, formally given by
\begin{align}
V^\dagger(x^-,\bxp) = P_{x^-} \exp{\left[ -i \int^{x^-}_{-\infty} dx'^- \partial^{-2}_\perp \rho_{\rm cov}(x'^-,\bxp) \right]}\ .\label{Eq:V}
\end{align}
Here, $\rho_{\rm cov}$ is the color charge density in the covariant gauge condition and is related to the color charge density in the $A_- = 0$ gauge condition through the gauge transformation,
\begin{align}
\rho_{\rm cov} = V^\dagger \rho_{_{\rm (A_-=0)}} V\ .\label{Eq:gt}
\end{align}
As shown in Eq.~\refeq{Eq:A_single} and Eq.~\refeq{Eq:V}, the solution to the equation of motion is a functional of $\rho$, and thus the event average of a given observable $\mathcal{O}$ is obtained as the ensemble average over $P[\rho]$,
\begin{align}
\average{\mathcal{O}}_{\rm eve} = \int \mathcal{D}\rho \mathcal{O}[\rho] P[\rho]\ \label{Eq:eveave}.
\end{align}
Fortunately, in the high-energy limit, $P[\rho]$ can be well approximated by the normal distribution function, no matter what gauge condition is chosen,
\begin{align}
P[\rho(x^-,\bxp)] \propto \exp{\left[-\frac{{\rm Tr}[\rho(x^-,\bxp)]^2}{2[g^2\mu(x^-,\bxp)]^2}\right]}\ .
\end{align}
where $[g^2\mu(x^-,\bxp)]^2$ is the squared color charge density per unit volume $dx^- dx dy$. Therefore, in the high-energy limit, the event average in Eq.\refeq{Eq:eveave} can be estimated numerically using the Gaussian random number $\rho_{\rm cov}$ that satisfies the following event average,
\begin{align}
&\average{ \rho^a_{\rm cov}(x^-,\bxp)\rho^b_{\rm cov}(x'^-,\bxp') }_{\rm eve}\nonumber\\
&= \delta^{a,b} \left( g^2\mu(x^-,\bxp) \right)^2 \delta(x^--x'^-) \delta^2(\bxp-\bxp')\ .\label{Eq:mv_rho}
\end{align}
It should be mentioned that solving the JIMWLK equation~\cite{JIMWLK1997,JIMWLK1998,JIMWLK2001,JIMWLK2002,JIMWLK_1_2001,JIMWLK_2_2001,JIMWLK_LM}, the evolution equation for momentum rapidity, yields the Wilson line at the energy of interest beyond the high-energy limit approximation.

Using the MV model with the high-energy limit approximation, the glasma created in the collision of two nuclei can be obtained as a boost-invariant CYM field.
In this approximation, hard partons are assumed to be recoilless, and thus the total classical color current is given by the incoherent sum of the two static color currents,
\begin{align}
J^{\mu}(x) = \frac{1}{g} \delta^{\mu+} \delta(x^-) \rho^{(1)}(\bm{x}_\perp)
           + \frac{1}{g} \delta^{\mu-} \delta(x^+) \rho^{(2)}(\bm{x}_\perp)\ ,\label{Eq:J_double}
\end{align}
where the color charge density from each nucleus is assumed to be distributed on an infinitely thin sheet due to the Lorentz contraction, $\rho^{(1/2)}(x^\mp,\bxp) \propto \delta(x^\mp)$.
Then, solving the classical equation of motion $[D_\mu, F^{\mu\nu}]=J^\nu$ with the Fock-Schwinger (FS) gauge condition $A^\tau=0$ yields the initial condition of the glasma at $\tau=0^+$ as a regular solution,
\begin{align}
A_i&= A^{(1)}_i + A^{(2)}_i\ ,\ \ A_\eta=0\ ,\label{Eq:2Dsolution_A}\\
E^i&= 0\ ,\ \ E^\eta=ig[A^{(1)}_i, A^{(2)}_i]\ ,\label{Eq:2Dsolution_E}
\end{align}
where the transverse and longitudinal electric fields are defined as $E^i=\tau \partial_\tau A_i$ and $E^\eta= \partial_\tau A_\eta/\tau$, respectively, and $A^{(1/2)}_i$ is the transverse gauge field emitted from the single nucleus $1$ or $2$, respectively, 
\begin{align}
A^{(1/2)}_i = \frac{i}{g} V^{(1/2)}_{\rm 2D} \partial_i V^{(1/2)\dagger}_{\rm 2D}\ .
\end{align}
In this paper, the index $i$ denotes the transverse directions, $1$ and $2$, unless otherwise stated. The Wilson line $V^{(1/2)\dagger}_{\rm 2D}$ is independent of $x^\mp$ and is given by
\begin{align}
V^{(1/2)\dagger}_{\rm 2D}(\bxp) = P_{x^\mp} \exp{\left[ -i \int^{\infty}_{-\infty} dx'^\mp \partial^{-2}_\perp \rho^{(1/2)}_{\rm cov}(x'^\mp,\bxp) \right]}\ .\label{Eq:Wilson2D}
\end{align}
Here, since two classical color charges are only located on the light-cone, $\rho^{(1/2)} \propto \delta(x^\mp)$, the upper limit of the integration in Eq.~\refeq{Eq:Wilson2D} can be taken as infinity. Therefore, the solutions shown in Eq.\refeq{Eq:2Dsolution_A} and Eq.\refeq{Eq:2Dsolution_E} are boost invariant. To study the boost-invariant glasma at a late time, we have to evolve the CYM field starting from the boost-invariant initial condition by solving the classical equation of motion inside the light-cone ($J=0$), $[D_\mu, F^{\mu\nu}]=0$. 

\subsection{$3+1$D glasma in continuous spacetime}\label{Sec:3dcont}
\begin{figure}[htbp]
  \begin{center}
  \includegraphics[width=75mm,bb=0 0 600 352]{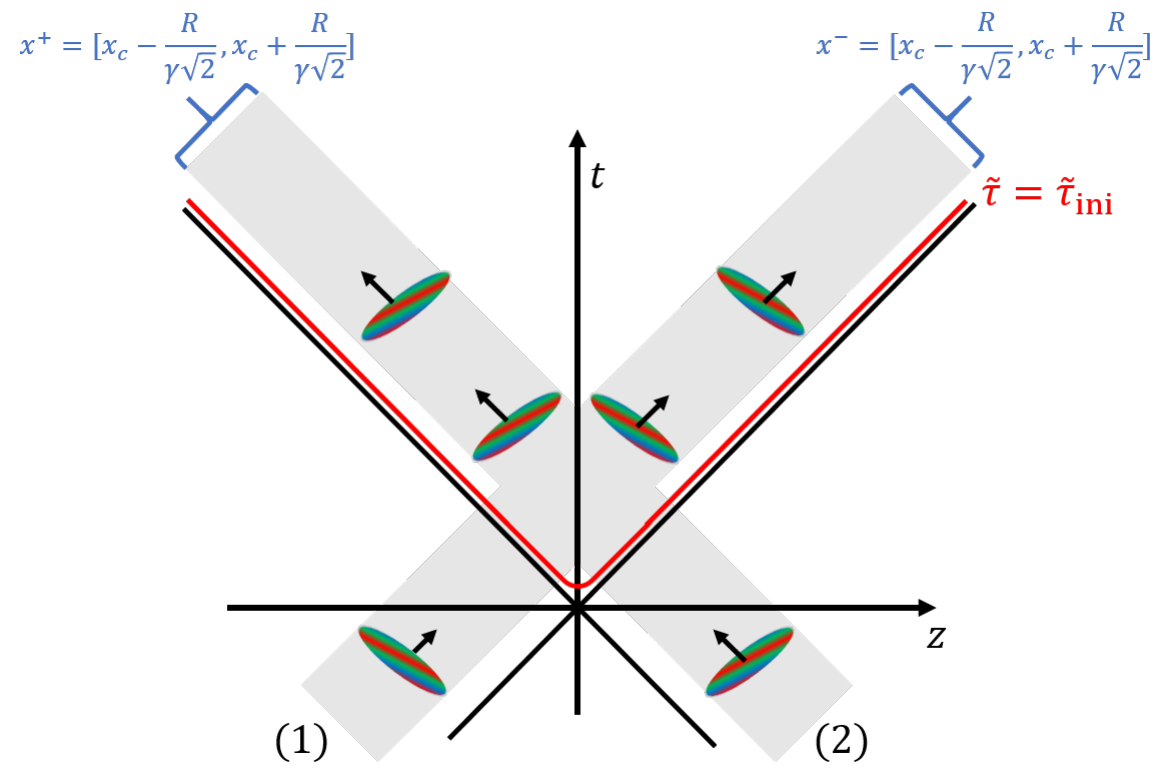}
  \end{center}
  \caption{Spacetime picture of a collision of relativistic nuclei with the finite longitudinal extension $R/\gamma$. At initial proper time $\ttau_{\rm ini}$, the two nuclei are still apart from each other.}
  \label{Fig:spacetime}
  \end{figure}
We explain how to extend the $2+1$D glasma description to the $3+1$D glasma description with the dynamical $3$D classical color current that represents two colliding nuclei with finite longitudinal thickness. As an example, we consider the situation where the two colliding nuclei have the same radius $R$ and Lorentz gamma factor $\gamma$. The generalization to the collisions of two nuclei with different radiuses and gamma factors can be carried out straightforwardly. Let us first revisit the total classical color current to get the initial condition for the $3+1$D glasma. Our $3+1$D glasma method considers the setup where the two nuclei are still far apart at the initial proper time $\ttau=\ttau_{\rm ini}$ in the Milne coordinate defined as $(\tilde{\tau}=\sqrt{2\xm\xp},\tilde{\eta}=(1/2)\ln({\xp}/{\xm}))$. It should be noted that we distinguish the Milne coordinates defined here from the usual Milne coordinates $(\tau=\sqrt{2(\xm - x_{\rm c})(\xp - x_{\rm c})},\eta=(1/2) \ln({[\xp - x_{\rm c}]}/{[\xm - x_{\rm c}]})))$ in which central positions of the two nuclei coincide at $\tau=0$. The center positions of the nuclei $(1)$ and $(2)$ in the Milne coordinates are initially taken as a sufficiently large negative and positive value, $-|\teta_{\rm ini}|$ and $|\teta_{\rm ini}|$. The corresponding center positions in the light-cone coordinates are given by $x^\mp=x_{\rm c}=\ttau_{\rm ini} e^{|\teta_{\rm ini}|}/\sqrt{2}$, and thus the nuclei $(1)$ and $(2)$ exist within $\xm=[x_{\rm c}-R/(\gamma\sqrt{2}), x_{\rm c}+R/(\gamma\sqrt{2})]$ and $\xp=[x_{\rm c}-R/(\gamma\sqrt{2}), x_{\rm c}+R/(\gamma\sqrt{2})]$, respectively, as shown in Fig.~\ref{Fig:spacetime}. The initial proper time $\tau_{\rm ini}$ is set so small that the two nuclei do not overlap at $\ttau=\ttau_{\rm ini}$, which requires the relation, $x^\mp|_{\teta=0,\ttau=\ttau_{\rm ini}} = \tau_{\rm ini}/\sqrt{2} < x_{\rm c}-R/(\gamma\sqrt{2})$. Then, the classical color current at $\tilde{\tau}_{\rm ini}$ can be assumed to be the incoherent sum of the two classical color currents,
\begin{align}
J^{\mu}(x) = \frac{1}{g} \delta^{\mu+} \rho^{(1)}(x^-,\bm{x}_\perp)
           + \frac{1}{g} \delta^{\mu-} \rho^{(2)}(x^+,\bm{x}_\perp)\ .\label{Eq:J_double_3d}
\end{align}
The transverse gauge field and electric field at $\tilde{\tau}=\tilde{\tau}_{\rm ini}$ are also assumed to be the incoherent sum of those from each nucleus,
\begin{align}
A_i(x) &=                A^{(1)}_i(x) +               A^{(2)}_i(x)\ ,\\
E^i(x) &=                E^{(1)i}(x)  +               E^{(2)i}(x) \nonumber\\
       &= x^- \partial_- A^{(1)}_i(x) + x^+\partial_+ A^{(2)}_i(x)\ ,\label{Eq:CYMini1}
\end{align}
where the transverse gauge field from a single nucleus, $A^{(1/2)}_i(x)$, is given by the Wilson line as given in Eq.~\refeq{Eq:V}, and the electric gauge field, $E^{(1/2)i}=x^\mp \partial_\mp A^{(1/2)}_i$, is obtained by using the relation $\tau \partial_\tau = \xm \partial_- + \xp \partial_+$. The longitudinal components of the gauge field and electric field are given in the same form as those shown in Eq.\refeq{Eq:2Dsolution_A} and Eq.\refeq{Eq:2Dsolution_E},
\begin{align}
A_{\tilde{\eta}} = 0\ ,\ \ \ E^{\tilde{\eta}} = ig[A^{(1)}_i, A^{(2)}_i]\ .\label{Eq:CYMini2}
\end{align}
We have used the modified FS gauge condition $A^{\tilde{\tau}}=0$, and then this set of initial conditions satisfies Gauss's law $[D_\mu,F^{\mu\ttau}]=J^{\ttau}$.
It should be noted that $E^{\tilde{\eta}}$ is negligibly small since the two nuclei are largely apart from each other at $\tilde{\tau}=\tilde{\tau}_{\rm ini}$.

In the actual calculation, we present the initial condition given above on a lattice and evolve them by solving their evolution equations numerically. 
The evolution equation for the CYM field is the classical equation of motion $[D_\mu, F^{\mu\nu}]=J^\nu$ and the evolution equation for the classical current is the continuity equation $[D_\mu,J^\mu]=0$. Since $J^\mu$ has two degrees of feedom as $J^\mu=\delta^{\mu +} J^{(1)}+\delta^{\mu -} J^{(2)}$, 
an additional assumption is required so that $[D_\mu, F^{\mu\nu}]=J^\nu$ and $[D_\mu, J^\mu]=0$ form a closed system of equations.
In this study, we assume that $J^{(1/2)}$ obeys the continuity equation for each nucleus, $[D_\pm, J^{(1/2)}] = 0$. 
This assumption is valid if at least two nuclei do not overlap, e.g., before the collision or after the two nuclei have passed.
To more accurately track the evolution of the color current beyond this assumption, it is necessary to solve the equations of motion for the microscopic degrees of freedom that carry the color charge.
For convenience, we introduce the current defined as $\tilde{J}^{(1/2)} \equiv x^\mp J^{(1/2)}$ and rewrite the continuity equation in the Milne coordinates as,
\begin{align}
(\tilde{\tau}\partial_{\tilde{\tau}} \pm D_{\tilde{\eta}} ) \tilde{J}^{(1/2)}=0\ .\label{Eq:cont_J12}
\end{align}
The classical equation of motion and Gauss's law in the Milne coordinates are written as
\begin{align}
\partial_{\ttau} E^1       &= 
-         \ttau  [D_2,B^3] 
+\frac{1}{\ttau} [D_3,B^2]\ ,\\
\partial_{\ttau} E^2       &= 
          \ttau  [D_1,B^3] 
-\frac{1}{\ttau} [D_3,B^1]\ ,\\
\partial_{\ttau} E^{\teta} &= 
- \frac{1}{\ttau} \epsilon^{\teta jk} [D_j,B^k] 
- \frac{1}{\ttau}\left[\tJo - \tJt\right]\ ,\label{Eq:eom_j}\\
[D_i,E^i]          &= \tJo + \tJt\ ,\label{Eq:gw_j}
\end{align}
where $\epsilon_{\teta jk}$ is Levi-Civita symbol, and the magnetic field $B$ is defined as $B^i=\epsilon^{ijk} F_{jk}/2\ \ (i=1,2,\teta)$.

\subsection{$3+1$D glasma in discretized space and continuous proper time}\label{Sec:3discr}
For numerical calculations, we discretize the CYM field and classical current on the $L^2_\perp \times L_{\tilde{\eta}}$ lattice, whose grid positions are labeled by a set of integers, $(i_x=0,1,\cdots,L_\perp-1, i_y=0,1,\cdots,L_\perp-1, i_{\tilde{\eta}}=0,1,\cdots,L_{\tilde{\eta}}-1)$.
These integers are related to spatial coordinates as $x=a_\perp(i_x-(L_\perp-1)/2)$, $y=a_\perp(i_y-(L_\perp-1)/2)$ and $\tilde{\eta} = a_{\tilde{\eta}}(i_{\tilde{\eta}} - L_{\tilde{\eta}}/2)$ with the lattice spacings, $a_\perp$ and $a_{\teta}$.
All the quantities shown in this and later sections are made dimensionless normalizing with the transverse spatial lattice spacing $a_\perp$, and the $\teta$ component of the gauge field $A_{\teta}$ is normalized by the longitudinal lattice spacing $a_{\teta}$.

We first consider the initial condition, the equation of motion, and the Gauss's law for the discretized CYM field.
The gauge field and longitudinal electric field at the initial proper time, shown in Eq.~\refeq{Eq:CYMini1} and Eq.~\refeq{Eq:CYMini2}, are discretized in a way that has been done in many papers (first in Ref.~\cite{2Dglasma_KV}),
\begin{align}
U_{i,x}     &= \left( U^{(1)}_{i,x} + U^{(2)}_{i,x} \right)\left( U^{(1)\dagger}_{i,x} + U^{(2)\dagger}_{i,x} \right)^{-1}\ ,\ \ \ U_{\teta,x} &= I\ ,
\end{align}
and
\begin{align}
E^{\tilde{\eta}}_x&=\frac{i}{4 g} \sum_{i} 
\Bigl[ 
  \left( U_{i,x+\hat{\teta}/2}                 - I \right) \left( U^{(2)\dagger}_{i,x+\hat{\teta}/2}  - U^{(1)\dagger}_{i,x+\hat{\teta}/2}  \right)\nonumber\\
&+\left( U^\dagger_{i,x+\hat{\teta}/2-\hat{i}} - I \right) \left( U^{(2)}_{i,x+\hat{\teta}/2-\hat{i}} - U^{(1)}_{i,x+\hat{\teta}/2-\hat{i}} \right)
 -{\rm \ h.c.}
  \Bigr]\ ,
\end{align}
where $U^{(1/2)}_{i,x}= V^{(1/2)}_x V^{(1/2)\dagger}_{x+\hat{i}}$ is the link variable for nucleus $1/2$, respectively, and the way of evaluating the Wilson line on the lattice is given in Appendix~\ref{App:Wilson}. The initial transverse electric field on the lattice is given by
\begin{align}
E^i_x 
&= \frac{a_{\tilde{\eta}} \xm}{g} V^{(1)}_x \left[ \partial^{\rm F}_i \partial^{-2}_\perp \rho^{(1)}_{\rm cov}(x^-,\bxp) \right] V^{(1)\dagger}_x\nonumber\\
&+ \frac{a_{\tilde{\eta}} \xp}{g} V^{(2)}_x \left[ \partial^{\rm F}_i \partial^{-2}_\perp \rho^{(2)}_{\rm cov}(x^+,\bxp) \right] V^{(2)\dagger}_x\ ,
\end{align}
where $\partial^{\rm F}$ is a forward difference, and $\partial^{-2}_\perp \rho^{(1/2)}_{\rm cov}$ is obtained through discretized Fourier transformation in the transverse directions,
\begin{align}
\partial^{-2}_\perp \rho^{(1/2)}_{\rm cov}(x^\mp,\bxp) 
= \frac{1}{L^2_\perp} \sum^{L_\perp}_{k_1,k_2=0} \frac{\tilde{\rho}^{(1/2)}_{\rm cov}(x^\mp,\bm{k}_\perp)}{k^2_{{\rm lat},\perp}} e^{i\bx_\perp\cdot\bk_\perp}\ .\label{Eq:rho_fourier}
\end{align}
where $\bm{k}_\perp=(k_1,k_2)=2\pi/L(n_1,n_2)\ (n_1,n_2=0,1,\cdots,L_\perp-1)$ is a wave number on the lattice, $k_{{\rm lat},\perp}=2\sqrt{\sin^2\frac{k_1}{2}+\sin^2\frac{k_2}{2}}$ is the transverse momentum on the lattice, 
and $\tilde{\rho}^{(1/2)}_{\rm cov}$ 
is the discrete Fourier transform of $\rho^{(1/2)}_{\rm cov}$ in the transverse direction, 
$\tilde{\rho}^{(1/2)}_{\rm cov}(x^\mp,\bm{k}_\perp) = \sum_{x^1,x^2} \rho^{(1/2)}_{\rm cov}(x^\mp,\bxp) e^{-i\bx_\perp\cdot\bk_\perp}$. As will be explained in the later sections, we introduce the infrared regulator in $\rho$, and as a result, the classical color charge vanishes at $k_{{\rm lat},\perp}=0$. To obtain the equation of motion and Gauss's law for the discretized CYM field with the dynamical current, we begin with the case in the absence of the dynamical current~\cite{InstabilityCYM2009},
\begin{align}
&\partial_{\tilde{\tau}} U_{i,x}   = ig \frac{g_{ii} E^i_x}{a_{\teta} \ttau} U_{i,x}\ ,\\
&\partial_{\tilde{\tau}} E^i_x = - \frac{i a_{\tilde{\eta}}\tilde{\tau}}{2g }\sum_{j \neq i} g^{ii} g^{jj} 
\left[W_{ij,x} -  U^\dagger_{j,x-\hat{j}}W_{ij,x-\hat{j}}U_{j,x-\hat{j}} \right]\ ,\label{Eq:eom_e3d}\\
&\sum_{i=1,2,\tilde{\eta}} \left( E^i_x - U^\dagger_{i,x-\hat{i}}E^i_{x-\hat{i}}U_{i,x-\hat{i}} \right)=0\ ,\label{Eq:gw_3d}
\end{align}
where the index $i$ runs over $1,2$ and $\teta$, and $g^{\mu\nu}={\rm diag}(1,-1,-1,-(a_{\tilde{\eta}}\tilde{\tau})^{-2})$ is the metric of the Milne coordinates on the lattice, and $W_{ij,x}\equiv U_{ij,x} - U^\dagger_{ij,x}$ is the difference of the plaquette, $U_{ij,x}=U_{i,x}U_{j,x+\hat{i}}U^\dagger_{i,x+\hat{j}}U^\dagger_{j,x}$, and its Hermite conjugate.

In below we discuss the definition of the electric field on the lattice.
There are two ways to define the electric field on the lattice, the left and right electric fields, connected by the relation $E^i_{{\rm R},x}=-U^\dagger_{i,x-\hat{i}}E^i_{{\rm L},x-\hat{i}}U_{i,x-\hat{i}}$. The electric field we consider here is the left one, $E^i_x=E^i_{{\rm L},x}$. Therefore, we have to add the left current,  $\tilde{J}^{(1/2)}_{{\rm L}, x}$, to the equation of motion for $E^{\tilde{\eta}}$ as shown in Eq.~\refeq{Eq:eom_j},
\begin{align}
\partial_{\tilde{\tau}} E^{\teta}_x &= - \frac{i}{2g a_{\teta} \ttau}\sum_{j \neq \teta}
\left[W_{\teta j,x} -  U^\dagger_{j,x-\hat{j}}W_{\teta j,x-\hat{j}}U_{j,x-\hat{j}} \right]\nonumber\\
&- \frac{1}{\tilde{\tau}}\left[ \tJo_{{\rm L}, x} - \tJt_{{\rm L}, x} \right]\ .
\end{align}
where $\tilde{J}^{(1/2)}_{{\rm L}, x}$ is located on $x+\hat{\tilde{\eta}}/2$ as well as the left electric field $E^{\tilde{\eta}}_x$.
On the other hand, the Gauss's law, which is shown in Eq.~\refeq{Eq:gw_3d}, is independent of the choice of the electric field on the lattice since the left-hand side of Eq.~\refeq{Eq:gw_3d} is nothing but the sum of the left and right electric fields,
\begin{align}
\sum_{i=1,2,\tilde{\eta}} \left( E^i_x            - U^\dagger_{i,x-\hat{i}}E^i_{x-\hat{i}}U_{i,x-\hat{i}} \right)=
\sum_{i=1,2,\tilde{\eta}} \left( E^i_{{\rm L}, x} + E^i_{{\rm R}, x}                                      \right)\ .
\end{align}
Therefore, we add the current $\tilde{J}^{(1/2)}$, which is independent of the left/right choice, to Gauss's law as shown in Eq.\refeq{Eq:gw_j},
\begin{align}
\sum_{i=1,2,\tilde{\eta}} \left( E^i_x - U^\dagger_{i,x-\hat{i}}E^i_{x-\hat{i}}U_{i,x-\hat{i}} \right) = a_{\tilde{\eta}} \left( \tilde{J}^{(1)}_x + \tilde{J}^{(2)}_x \right)\ .\label{Eq:gw_j_lat}
\end{align}
Next, we consider the initial condition for the discretized currents $\tilde{J}^{(1/2)}_x$ and $\tilde{J}^{(1/2)}_{{\rm L},x}$, and the continuity equations for them.
The initial condition for $\tilde{J}^{(1/2)}$ in the modified Fock-Schwinger gauge condition, shown in Eq.~\refeq{Eq:J_double_3d}, is discretized as
\begin{align}
\tilde{J}^{(1/2)}_x = \frac{x^\mp}{g} V^{(1/2)}_x \rho^{(1/2)}_{{\rm cov},x} V^{(1/2)\dagger}_{x}\ .\label{Eq:init_J}
\end{align}
Here we employ the gauge transformations for the color charge density shown in Eq.~\refeq{Eq:gt}.
The initial condition for the left current is assumed to have the same expression as that for $\tilde{J}^{(1/2)}_x$,
\begin{align}
\tilde{J}^{(1/2)}_{{\rm L}, x} = \frac{x^\mp}{g} V^{(1/2)}_{x+\frac{\hat{\teta}}{2}} \rho^{(1/2)}_{{\rm cov},x+\frac{\hat{\teta}}{2}} V^{(1/2)\dagger}_{x+\frac{\hat{\teta}}{2}}\ .\label{Eq:init_JL}
\end{align}
Since $U_{\tilde{\eta}}=I$ at $\tilde{\tau}_{\rm ini}$, the longitudinal electric field $E^{\teta}$ is initially independent of the left/right choice, $E^{\teta}_{\rm L}=-E^{\teta}_{\rm R}$. 
Thus, it is reasonable to assume that the color current is also independent of the left/right choice at the initial proper time.
Then, to get the continuity equations, we perform $\tilde{\tau}$ derivative on the left and right hands of the Gauss's law given in Eq.~\refeq{Eq:gw_j_lat}, 
\begin{align}
&a_{\tilde{\eta}} \partial_{\tilde{\tau}} \left( \tilde{J}^{(1)}_x + \tilde{J}^{(2)}_x \right)
= - \frac{1}{\tilde{\tau}}\left[ \tilde{J}^{(1)}_{{\rm L},x} - \tilde{J}^{(2)}_{{\rm L},x} \right] \nonumber\\
& + \frac{1}{\tilde{\tau}} U^\dagger_{\tilde{\eta},x-\hat{\tilde{\eta}}}
    \left[ \tilde{J}^{(1)}_{{\rm L},x-\hat{\tilde{\eta}}} - \tilde{J}^{(2)}_{{\rm L},x-\hat{\tilde{\eta}}} \right]
    U_{\tilde{\eta},x-\hat{\tilde{\eta}}}\ .\label{Eq:der_gau}
\end{align}
In accordance with the discussions in the continuum limit (see discussions above Eq.~\refeq{Eq:cont_J12}), we assume that the color currents, $\tilde{J}^{(1)}$ and $\tilde{J}^{(2)}$, evolve according to
\begin{align}
&\tilde{\tau} \partial_{\tilde{\tau}} \tilde{J}^{(1/2)}_x =\mp\frac{1}{a_{\tilde{\eta}}}
\left[ \tilde{J}^{(1/2)}_{{\rm L},x} - U^\dagger_{\tilde{\eta},x-\hat{\teta}} \tilde{J}^{(1/2)}_{{\rm L},x-\hat{\teta}} U_{\teta,x-\hat{\teta}} \right]\ .\label{Eq:j_disc}
\end{align}
This equation agrees with Eq.~\refeq{Eq:cont_J12} in the continuum limit. In addition, following Ref.~\cite{3DglasmaSS}, we assume that the evolution equation for the left current is given as,
\begin{align}
\ttau \partial_{\ttau} \tilde{J}^{(1/2)}_{{\rm L},x} =\mp\frac{1}{a_{\teta}}\left[ U_{\teta,x} \tilde{J}^{(1/2)}_{x+\hat{\teta}} U^\dagger_{\teta,x} - \tilde{J}^{(1/2)}_x \right]\ .\label{Eq:jL_disc}
\end{align}
This equation also agrees with Eq.~\refeq{Eq:cont_J12} in the continuum limit.

\subsection{Energy-momentum tensor in discretized space}\label{Sec:EMtensor}
We define the energy-momentum (EM) tensor of the CYM field on the lattice.
In principle, the EM tensor on discrete spacetime cannot be defined as Noether current due to translational symmetry breaking by the lattice.
To define an appropriate ``EM tensor" on the grid point for the real-time lattice simulation, we translate the expression of the EM tensor in continuous spacetime onto lattice,
\begin{align}
T^{\mu\nu}_x &= -g^{\kappa\sigma} F_{({\rm grid})\mu\kappa,x} F_{({\rm grid})\nu\sigma,x}\nonumber\\
             &+ \frac{1}{4} g_{\mu\nu} g^{\alpha\beta} g^{\gamma\omega} F_{({\rm grid})\alpha\gamma,x} F_{({\rm grid})\beta\omega,x}\ ,
\end{align}
where the field strength on the grid point can be written with the electric and magnetic field on the grid point, 
\begin{align}
F^{i\ttau}_{({\rm grid})x} &= \frac{E^i_{({\rm grid})x}}{\sqrt{-{\rm det}g_{\mu\nu}}}\ ,\\
F^{ij}_{({\rm grid})x}     &= \epsilon^{ijk}B^k_{({\rm grid})x}\ .
\end{align}
The electric field on the grid point is defined as the distance between the left and right electric field,
\begin{align}
E^i_{({\rm grid})x} \equiv \frac{1}{2} \left[ E^i_{{\rm L},x} - E^i_{{\rm R},x+\hat{i}} \right]
= \frac{1}{2} \left[ E^i_x + U^\dagger_{\hat{i},x}E^i_x U_{i,x} \right]\ ,
\end{align}
and the magnetic field on the grid point is defined using the $4$ plaquettes in the neighborhood of the grid point,
\begin{align}
B^i_{({\rm grid})x} \equiv \frac{1}{8}
\Bigl[ 
\Bigl(                                     
 &U_{i,x}                 U_{j,x+\hat{i}}                 U^\dagger_{i,x+\hat{j}} U^\dagger_{j,x} \nonumber\\
+&U^\dagger_{j,x-\hat{j}} U_{i,x-\hat{j}}                 U_{j,x+\hat{i}-\hat{j}} U^\dagger_{i,x} \nonumber\\
+&U^\dagger_{i,x-\hat{i}} U^\dagger_{j,x-\hat{i}-\hat{j}} U_{i,x-\hat{i}-\hat{j}} U_{j,x-\hat{j}} \nonumber\\
+&U_{j,x}                 U^\dagger_{i,x-\hat{i}+\hat{j}} U^\dagger_{j,x-\hat{i}} U_{i,x-\hat{i}} 
\Bigr)
- ({\rm h.c.}) 
\Bigr]\ .
\end{align}
This discretized EM tensor should agree with the continuous one in the continuum limit.
To confirm that the $3+1$D glasma simulation performed on the lattice works without problems, in the following section, we check the two continuity equations. After that we will apply our method the the central Au-Au collisions.

\section{Numerical Results}\label{Sec:result}
We show the numerical results of the $3+1$D glasma simulations using the SU($2$) CYM field in the Milne coordinates. Our numerical simulations are performed on the $L^2_\perp \times L_{\teta}$ lattice. The discretized evolution equations on the lattice are given in Appendix~\ref{App:evo} and solved with the leap-flog method. The boundary condition in the transverse directions is periodic, and the CYM field and classical color current are imposed to vanish on the boundary of the $\teta$ direction. This section is organized as follows: In Sec.~\ref{Sec:result1}, we test two relations that should hold in continuous spacetime and check the consistency with calculations performed in Ref.~\cite{3DglasmaSS}. In Sec.~\ref{Sec:result2}, we show the evolution of some observables using the setup that corresponds to the central collisions of Au-Au at $\sqrt{s}=200$ GeV.

\subsection{Check of our calculations}\label{Sec:result1}
We first confirm that our simulations do not violate two relations derived from the continuity equation. Then, we calculate the transverse pressure and energy density in the local rest frame and check their consistency with those calculated in Ref.~\cite{3DglasmaSS}. The paper~\cite{3DglasmaSS} simulates the $3+1$D glasma evolution on a lattice using the Minkowski coordinates.

The initial color charge density considered here is assumed to be the multiplication of the $1$-dimensional normal distribution function $N_{\rm 1D}$ with the variance $R/(\gamma\sqrt{2})$, which represents the longitudinal shape of the nucleus, and the random number $\Gamma^{(1/2)}$
\begin{align}
\rho^{(1/2)}(x^\mp,\bxp) = N_{\rm 1D}(x^\mp-x_{\rm c}, \frac{R}{\gamma\sqrt{2}}) \Gamma^{(1/2)}(\bxp)\ ,
\end{align}
where $x_{\rm c}$ is the center position of the nuclei in the $x^\mp$ direction, and $R$ and $\gamma$ are the radius and the gamma factor of the nuclei.
The random number $\Gamma^{(1/2)}$ satisfies the following event average,
\begin{align}
\average{ \Gamma^{(1/2)a}(\bxp) \Gamma^{(1/2)b}(\bxp') }_{\rm eve} = \delta^{a,b} Q^2_s N_{\rm 2D}(\bxp-\bxp', \sigma_\perp)\ ,\label{Eq:Gamma}
\end{align}
where $Q_s$ is the saturation scale.
To introduce an ultraviolet cutoff for the transverse momentum of $\rho$, we use the $2$-dimensional normal distribution function with the variance $\sigma_\perp$, $N_{\rm 2D}(\bxp-\bxp', \sigma_\perp)$, in Eq.~\refeq{Eq:Gamma} instead of the delta function shown in Eq.~\refeq{Eq:mv_rho}.
The transverse ultraviolet cutoff is necessary to regulate divergence in the local operator of the gauge fields~\cite{div1,div2}.
In addition, we also introduce an infrared cutoff $m$ by multiplying the regulation factor by the color charge density in the transverse momentum space,
\begin{align}
\tilde{\rho}^{(1/2)}(x^\mp,\bm{k}_\perp) \to \frac{k^2_{{\rm lat},\perp}}{m^2+k^2_{{\rm lat},\perp}} \tilde{\rho}^{(1/2)}(x^\mp,\bm{k}_\perp)\ ,
\end{align}
which means that the contribution of scale less than $m$ is suppressed.

In the calculations in this section, the parameters shown in Table ~\ref{Tab1} are used, which is consistent with the previous calculations in Ref.~\cite{3DglasmaSS}.
The system size in the longitudinal direction, $a_{\teta} \times L_{\teta}$, is taken such that both nuclei are included in the lattice. The lattice spacing in the longitudinal direction, $a_{\teta}$, is small enough not to affect the results. The center position of the nuclei in the light-cone coordinates, $x^\mp=x_{\rm c}$, is taken such that the overlap of the two incoming nuclei is negligibly small at the initial proper time, $\ttau_{\rm ini}$.
\begin{table}[htbp]
\caption{Parameters used in Sec.~\ref{Sec:result1}}
\label{Tab1}
\begin{center}
\begin{tabular}{r|c}
\hline\hline
$L_\perp$         & $128$                                     \\
$L_{\teta}$       & $224, 448, 896$                           \\
$a_\perp$         & $1/(8Q_s)$                                \\
$a_{\teta}$       & $10/L_{\teta}$                            \\
$m$               & $Q_s$                                     \\
$\sigma_\perp$    & $\sqrt{2}/(10Q_s)$                          \\
$R/\gamma$        & $1/(2Q_s), 1/(4Q_s), 1/(8Q_s), 1/(16Q_s)$ \\
$\ttau_{\rm ini}$ & $0.1 a_\perp$                               \\
$x_{\rm c}$       & $\ttau_{\rm ini}/\sqrt{2} + 3R/\gamma$      \\
\hline\hline  
\end{tabular}
\end{center}
\end{table}
\begin{figure}[htbp]
\begin{center}
\includegraphics[width=75mm,bb=0 0 400 252]{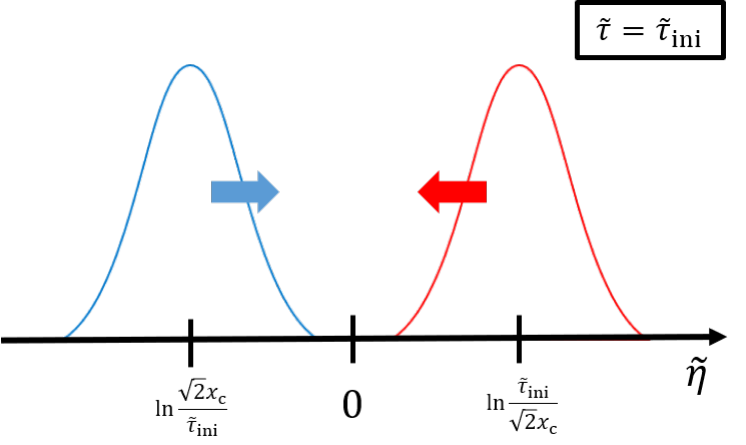}
\end{center}
\caption{Two incoming nuclei at the initial proper time $\ttau_{\rm ini}$.}
\label{Fig:initial}
\end{figure}

\begin{figure}[htbp]
\begin{center}
\includegraphics[width=0.9\linewidth]{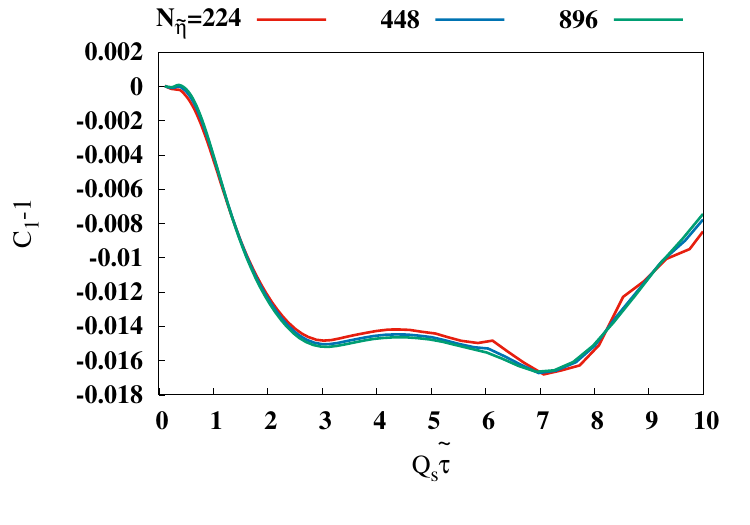}
\includegraphics[width=0.9\linewidth]{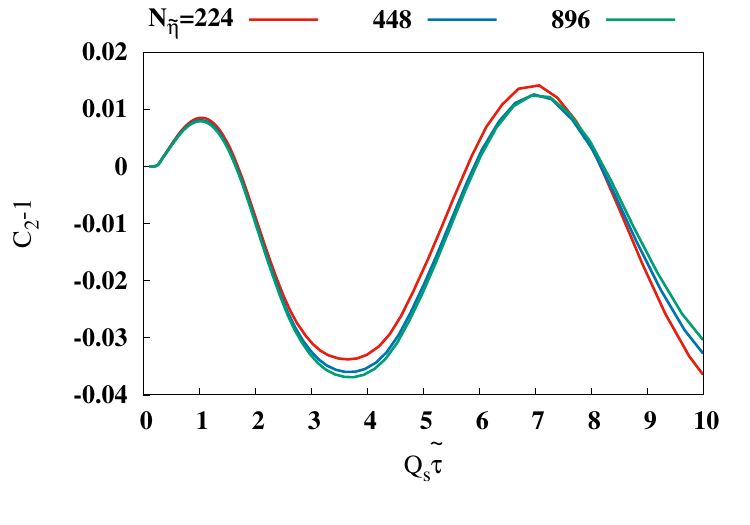}
\end{center}
\caption{
Test of the violation of relations~\refeq{Eq:ce1} and \refeq{Eq:ce2}, which should hold in the continuum limit as a result of the continuity equations for the EM tensor. The upper and Lower panels show the quantities defined in Eq.~\refeq{Eq:c1} and Eq.~\refeq{Eq:c2} that are introduced to measure the violations of Eq.~\refeq{Eq:ce1} and Eq.~\refeq{Eq:ce2}, respectively. }
\label{Fig:c12}
\end{figure}

Here we consider the continuity equations in the Milne coordinates,
\begin{align}
&[D_\mu, T^{\mu\ttau}]=-E^{\teta}J^{\teta}\ ,\\
&[D_\mu, T^{\mu\teta}]=0\ ,
\end{align}
or expanded explicitly as 
\begin{align}
&\frac{1}{\ttau}\left\{ \partial_{\ttau} \left[\ttau T^{\ttau\ttau}\right] + \ttau^2 T^{\teta\teta}\right\}+\partial_1 T^{1\ttau}+\partial_2 T^{2\ttau}+\partial_{\teta} T^{\teta\ttau}\nonumber\\&\qquad\qquad\qquad\qquad\qquad\qquad\qquad\quad=-E^{\teta}J^{\teta}\ ,\\
&\left(\partial_{\ttau} T^{\teta\ttau}+\frac{3T^{\teta\ttau}}{\ttau}\right) + \partial_1 T^{1\teta}+\partial_2 T^{2\teta}+\partial_{\teta} T^{\teta\teta}=0\ .
\end{align}
By integrating the left-hand and right-hand sides of these equations over space and dropping the surface terms, we obtain
\begin{align}
&\ttau\partial_{\ttau} \tau^{\ttau\ttau}=- \left( \tau^{\ttau\ttau} + \ttau^2 \tau^{\teta\teta} + \kappa \right)\ ,\label{Eq:ce1}\\
&\ttau\partial_{\ttau} \left( \ttau^3 \tau^{\teta\ttau} \right) = 0\ .\label{Eq:ce2}
\end{align}
where $\tau^{\mu\nu} \equiv \int dxdyd\teta T^{\mu\nu}/{V}$ and $\kappa \equiv \int dxdyd\teta E^{\teta}J^{\teta}/{V}$.
To measure the violation of these relations in actual simulations, we use the following quantities,
\begin{align}
C_1 &\equiv 
-\frac{2a_\theta\left( \tau^{\ttau\ttau} + \ttau^2 \tau^{\teta\teta} + \kappa \right)|_{\theta = \theta_{\rm ini}+n a_\theta}}{\tau^{\ttau\ttau} |_{\theta = \theta_{\rm ini}+(n+1)a_\theta}-\tau^{\ttau\ttau} |_{\theta = \theta_{\rm ini}+(n-1)a_\theta}}\ ,\label{Eq:c1}\\
C_2 &\equiv  
\frac{ \ttau^3 \tau^{\teta\ttau}|_{\theta = \theta_{\rm ini}+n a_\theta} }{ \ttau^3 \tau^{\teta\ttau}|_{\theta = \theta_{\rm ini} = \ln{\tau_{\rm ini}}} }\ ,\label{Eq:c2}
\end{align}
where $\theta=\ln{\ttau}$ is the time variable used for solving the evolution equations numerically by the difference method, $\theta_{\rm ini}=\ln{\ttau_{\rm ini}}$ is $\theta$ at the initial proper time, $n$ is the time step, and $a_\theta$ is the step size 
(See Appendix~\ref{App:evo} for details.).
Both quantities are normalized in such a way that they approach $1$ when the violations of the continuity equations are smaller. In the upper and lower panels of Fig.~\ref{Fig:c12}, we show the evolution of $C_1-1$ and $C_2-1$, respectively, calculated with $N_{\teta}=224$, $448$ and $896$ using the common random number $\Gamma^{(1/2)}$ from the same seed.
The deviations of $C_1-1$ and $C_2-1$ from $0$ are found to be small and stable in changes of $N_{\teta}$.
Thus, the effect of the discretization on the dynamics is considered tiny in our calculations with large $N_{\teta}$ and small $a_{\teta}$.

Next, we calculate the transverse pressure and energy density in the local rest frame on a $L^2_\perp \times L_{\tilde{\eta}}$=$128^2 \times 448$ lattice.
To focus only on the EM tensor that the glasma has, we define the subtracted EM tensor as 
\begin{align}
T^{\mu\nu}_{\rm sub} \equiv T^{\mu\nu} - T^{\mu\nu}_{(1)} - T^{\mu\nu}_{(2)}\ ,\label{Eq:subt}
\end{align}
where $T^{\mu\nu}$ is the total EM tensor and $T^{\mu\nu}_{(1/2)}$ is the EM tensor of the nucleus $(1/2)$, respectively.
To evaluate $T^{\mu\nu}_{(1/2)}$, we run two additional simulations in parallel, considering only one nucleus.
If the color charge densities change little with collision, the subtracted EM tensor $T^{\mu\nu}_{\rm sub}$ can be considered as the EM tensor of the glasma.
This subtraction method has been used in Ref.~\cite{3DglasmaSS} as well.
To check the consistency of our method with theirs, we calculate the subtracted transverse pressure and subtracted energy density in the local rest frame, averaged over the transverse plane, which are also calculated in Ref.~\cite{3DglasmaSS},
\begin{align}
P_\perp &\equiv \frac{\int d^2x_\perp \left[ T^{11}_{\rm sub}+T^{22}_{\rm sub} \right]}{2V_\perp}\ ,\\
\varepsilon_{\rm LRF}     &\equiv 
\frac{ \int d^2x_\perp
\Big[ T^{11}_{\rm sub}+T^{22}_{\rm sub} + \sqrt{\left(T^{\ttau\ttau}_{\rm sub} + T^{\teta\teta}_{\rm sub}\right)^2 - 4(T^{\teta\ttau}_{\rm sub})^2 }\Big]
}{2V_\perp}\ \label{Eq:pl_milne}
\end{align}
where $V_\perp=\int d^2{\bm x}_\perp$. In Ref.~\cite{3DglasmaSS}, these quantities are calculated using Minkovski coordinates, and thus the expression of the energy density in the local rest frame is different, 
\begin{align}
\varepsilon_{\rm LRF} \equiv 
\frac{ \int d^2x_\perp
\Big[ T^{11}_{\rm sub}+T^{22}_{\rm sub} + \sqrt{\left(T^{00}_{\rm sub} + T^{33}_{\rm sub}\right)^2 - 4\left(T^{30}_{\rm sub}\right)^2 }\Big]
}{2V_\perp}\ .\label{Eq:pl_min}
\end{align}
The consistency between Eq.~\refeq{Eq:pl_milne} and Eq.~\refeq{Eq:pl_min} can be checked using a general coordinate transformation.

\begin{figure}[htbp]
\begin{center}
\includegraphics[width=0.7\linewidth]{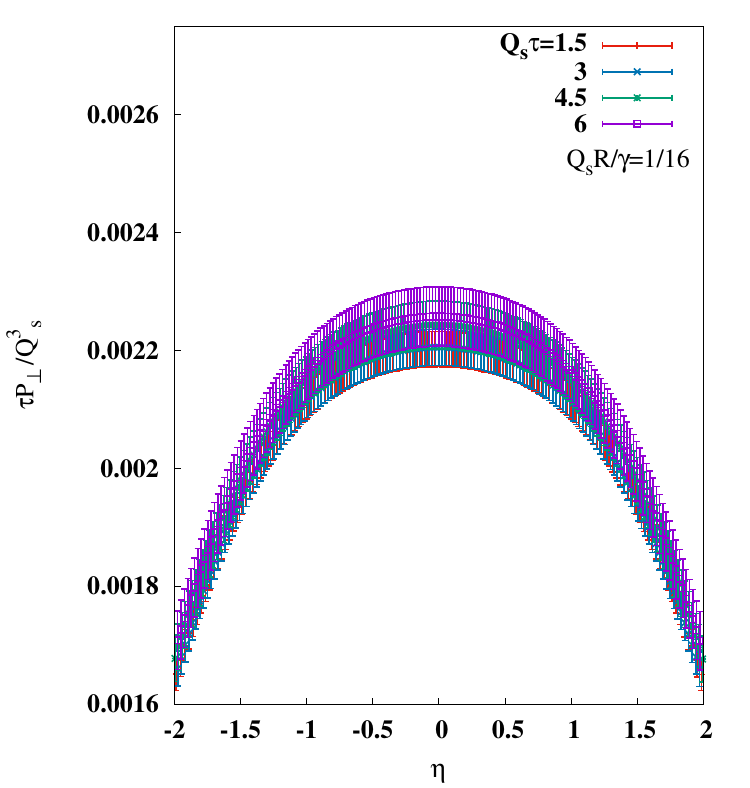}
\includegraphics[width=0.7\linewidth]{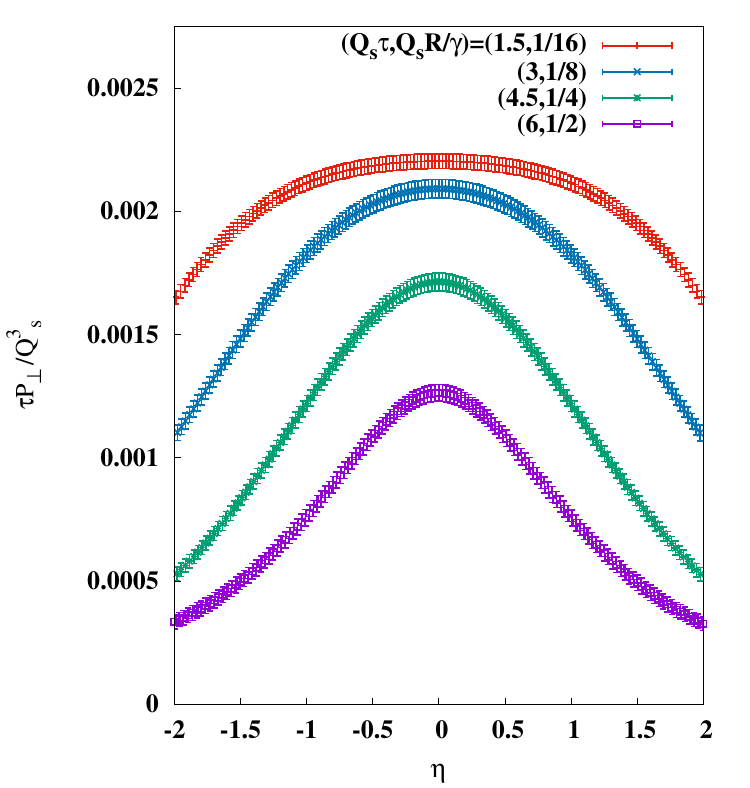}
\end{center}
\caption{
The $\eta$ dependence of the transverse pressure normalized by the proper time and the saturation scale, $\tau P_\perp/Q^3_s$.
All results shown here are event averages of $50$ independent simulations.
The upper panel shows $\tau P_\perp/Q^3_s$ for $Q_sR/\gamma=1/16$ at $Q_s \tau=1.5, 3.0, 4.5$ and $6.0$.
The lower panel shows $\tau P_\perp/Q^3_s$ for $Q_sR/\gamma=1/16, 1/8, 1/4$ and $1/2$, which are calculated at $Q_s \tau=1.5, 3.0, 4.5$ and $6.0$, respectively.}
\label{Fig:pt}
\end{figure}

Figure~\ref{Fig:pt} shows the $\eta$ dependence of the transverse pressures normalized by the proper time and the saturation scale, $\tau P_\perp/Q^3_s$, for different thicknesses, $Q_sR/\gamma=1/2, 1/4, 1/8$ and $1/16$.
These results are event averages of $50$ independent simulations, each given a different random number $\Gamma^{(1/2)}$.
Here $\tau=\sqrt{2(x^+-x_{\rm c})(x^--x_{\rm c})}$ and $\eta=\frac{1}{2}\ln{\frac{x^+-x_{\rm c}}{x^--x_{\rm c}}}$ is the usual Milne coordinates, in which central positions of the two nuclei coincide at $\tau=0$. Since the two Milne coordinates are in one-to-one correspondence as $(\tau=\tau(\ttau,\teta),\eta=\eta(\ttau,\teta))$, 
numerical simulations with discrete $(\ttau,\teta)$ can only provide observations at a large number of discrete points spread across the $(\tau,\eta)$ plane.
Therefore, the result at a fixed proper time $\tau$ shown in Fig.~\ref{Fig:pt} (and the following  figures) is actually sampled from results within $[0.99\tau,1.01\tau]$. The upper panel of Fig.~\ref{Fig:pt} shows that $\tau P_\perp/Q^3_s$ for $Q_sR/\gamma=1/16$ at $Q_s \tau=1.5, 3.0, 4.5$ and $6.0$ agree within the margin of error, which indicates that $P_\perp$ decreases as $\tau^{-1}$ at $1.5 \leq Q_s \tau \leq 6.0$.
While this scaling behavior is imposed as the assumption in the paper~\cite{3DglasmaSS}, we clarify that this scaling is established as time elapses since the collision. We present a discussion about the scaling behavior in Appendix~\ref{App:scaling}. The lower panel of Fig.~\ref{Fig:pt} shows results for different thicknesses at the late time, in which $P_\perp$ falls as $\tau^{-1}$.
It is found that the transverse pressures at different thicknesses have a similar peak around $\eta=0$ regardless of the thickness of the nucleus.
This peak around $\eta=0$ becomes milder as the nucleus becomes thinner.
This behavior is understandable since the glasma becomes boost-invariant when a nucleus is infinitely thin, as explained in Sec.~\ref{Sec:2d}.
These results reproduce well Fig.~$8$ in Ref.~\cite{3DglasmaSS}.
The most important point to note is that we can reproduce the these results using about $4.5$ times smaller number of grids in the longitudinal direction.
The number of grids in the $z$ direction in the calculation in Ref.~\cite{3DglasmaSS} is $2048$, while the number of grids in the $\teta$ direction in our calculation is $448$.

\begin{figure}[htbp]
\begin{center}
\includegraphics[width=0.7\linewidth]{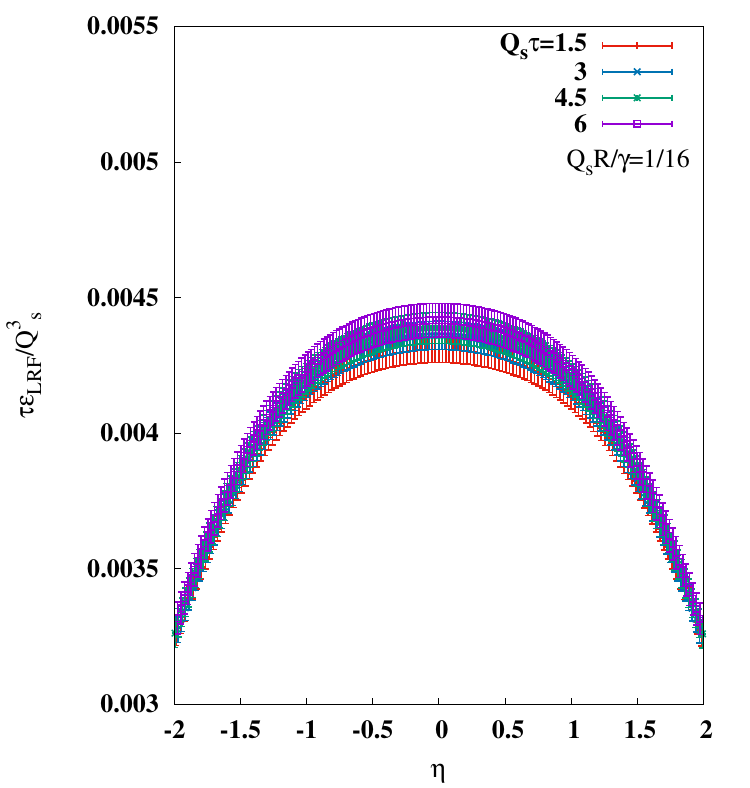}
\includegraphics[width=0.7\linewidth]{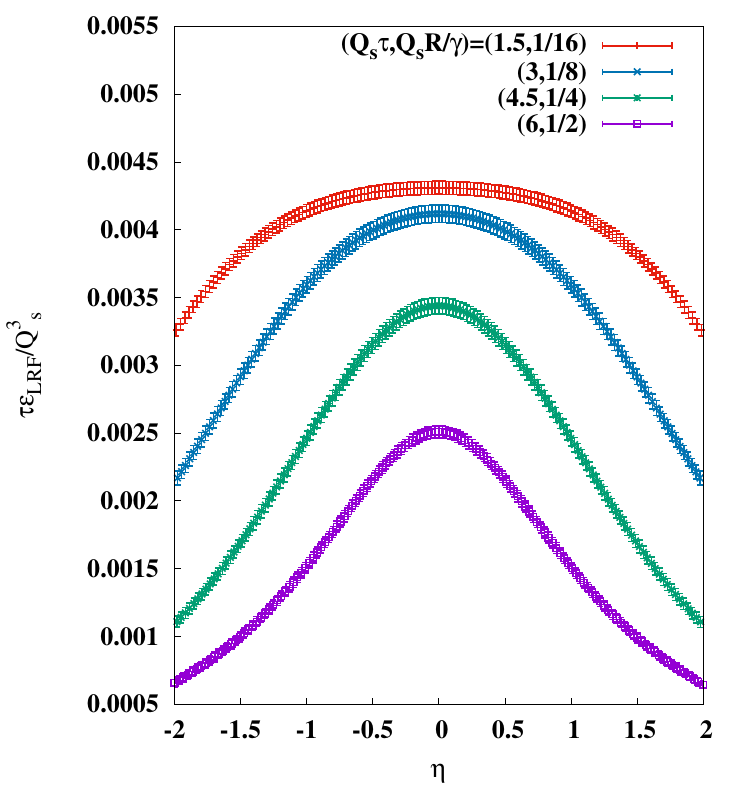}
\end{center}
\caption{
The $\eta$ dependence of the energy density in the local rest frame normalized by the proper time and the saturation scale, $\tau \varepsilon_{\rm LRF}/Q^3_s$.
The upper and lower panels of Fig.~\ref{Fig:e} are obtained from the same simulations as shown in the upper and lower panels of Fig.~\ref{Fig:pt}, respectively.}
\label{Fig:e}
\end{figure}

Figure~\ref{Fig:e} shows the $\eta$ dependence of the energy density in the local rest frame normalized by the proper time and the saturation scale, $\tau \varepsilon_{\rm LRF}/Q^3_s$.
The upper and lower panels of Fig.~\ref{Fig:e} are obtained from the same simulations as shown in the upper and lower panels of Fig.~\ref{Fig:pt}, respectively.
The upper panel of Fig.~\ref{Fig:e} shows that, in the late time region when $P_\perp$ decreases as $\tau^{-1}$, $\varepsilon_{\rm LRF}$ also decreases as $\tau^{-1}$.
The lower panel of Fig.~\ref{Fig:e} shows that $\varepsilon_{\rm LRF}$ is about $2$ times $P_\perp$ at the late time, which means that the transverse pressure $P_\perp$ is much larger than the longitudinal pressure in the local rest frame, defined as $P_{{\rm LRF},L} \equiv \varepsilon_{{\rm LRF}}-2P_\perp$. This definition of $P_{\rm L}$ is obtained by reference to the traceless of the EM tensor, $T^\mu_\mu=0$, which is a consequence of the conformal symmetry.
It must be mentioned here that the lower panel of Fig.~\ref{Fig:e} is inconsistent with the lower figure of Fig.~$9$ in Ref.~\cite{3DglasmaSS} and the discrepancy of $\varepsilon_{\rm LRF}$ between our and their results becomes larger as $Q_sR/\gamma$ becomes smaller. 
Given that our and their calculations for the transverse pressure completely agree, this discrepancy in local energy density is not understood and we leave the investigation of the cause of the discrepancy to a future task.
Since the continuity equations are not violated in our simulations, and our calculations are stable for varying lattice sizes and spacings, we believe that there is no fatal problem in our method, at least with respect to the initial condition and the dynamical evolution of the $3+1$D glasma.

\subsection{Central collisions of Au-Au}\label{Sec:result2}
We show the numerical results using the initial conditions that describe the central Au-Au collisions at $\sqrt{s}=200$ GeV. The color charge density at the initial proper time $\ttau_{\rm ini}$ is given as the incoherent sum of the color charge density of each nucleon.
The color charge density of $i-$th nucleon with a radius $R_{\rm n}$ is assumed to have the Gaussian shape whose center position is $(b^1_i,b^2_i,b^\mp_i)$,
\begin{align}
&\rho^{(1/2)}_i(x^\mp,\bxp)\nonumber\\
&= N_{\rm 1D}(x^\mp-b^\mp_i,\frac{R_{\rm n}}{\sqrt{6}\gamma})N_{\rm 2D}(\bxp-\bbpi,\frac{R_{\rm n}}{\sqrt{3}}) \Gamma^{(1/2)}_i(x^\mp,\bxp)\ .\label{Eq:rho_i}
\end{align}
The random number $\Gamma^{(1/2)}_i$ satisfies the following event average,
\begin{align}
&\average{ \Gamma^{(1/2)a}_i(x^\mp,\bxp) \Gamma^{(1/2)b}_i(x'^\mp,\bxp') }_{\rm eve} \nonumber\\
&= 
\delta^{a,b}
2\pi\left(\frac{2R^2_{\rm n}}{3}+\sigma^2_\perp\right)
\sqrt{ 2\pi\left(\frac{R^2_{\rm n}}{3\gamma^2}+\sigma^2_\mp\right) }
\left( g^2\bar{\mu} \right)^2 \nonumber\\
&\qquad\times N_{\rm 1D}(x^\mp - x'^\mp,\sigma_\mp)N_{\rm 2D}(\bxp-\bxp',\sigma_\perp)\ ,\label{Eq:gamma_i}
\end{align}
where $\sigma_\perp$ and $\sigma_\mp$ are the correlation lengths of $\Gamma^{(1/2)}$ in the transverse and longitudinal direction, respectively, and $g^2\bar{\mu}$ is the parameter controlling the strength of the color charge density.
Then, we can obtain the following relation (detailed derivation is presented in Appendix~\ref{App:rho}),
\begin{align}
&\average{ \rho^{(1/2)a}_i(x^\mp,\bxp)\rho^{(1/2)b}_i(x'^\mp,\bxp') }_{\rm eve} \nonumber\\
&= \delta^{a,b} \left( g^2\bar{\mu} \right)^2 
        N_{\rm 1D}(\frac{x^\mp+x'^\mp}{2}-b^\mp_i,\frac{R_{\rm n}}{\sqrt{3}\gamma}) N_{\rm 1D}(x^\mp - x'^\mp,l_\mp)\nonumber\\
&\times N_{\rm 2D}(\frac{\bxp+\bxp'}{2}-\bbpi,\sqrt{\frac{2}{3}}R_{\rm n})            N_{\rm 2D}(\bxp-\bxp',l_\perp)\ ,\label{Eq:rho_i_v2}
\end{align}
where $l_\perp$ and $l_\mp$ are the correlation lengths of $\rho^{(1/2)}$, which are related to $\sigma_\perp$ and $\sigma_\mp$ by
\begin{align}
l^{-2}_\perp &= \sigma^{-2}_\perp + \left(\sqrt{\frac{2}{3}} R_{\rm n}\right)^{-2}\ ,\label{Eq:lp}\\
l^{-2}_\mp   &= \sigma^{-2}_\mp   + \left(\frac{R_{\rm n}}{\sqrt{3}\gamma}\right)^{-2}  \ .\label{Eq:lmp}
\end{align}
We can define the squared color charge density of the nucleon per unit volume $dx^- dx dy$ as
\begin{align}
\left(g^2\mu(x^\mp,\bxp)\right)^2 \equiv \left( g^2\bar{\mu} \right)^2 N_{\rm 1D}(x^\mp,\frac{R_{\rm n}}{\sqrt{3}\gamma}) N_{\rm 2D}(\bxp,\sqrt{\frac{2}{3}}R_{\rm n})\ .
\end{align}
and its integration over $x^\mp$ at $\bxp=0$ is assumed to be proportional to the nucleon saturation scale, $g^2\mu_{\rm 2D, c}=\int dx^\mp g^2\mu(x^\mp,0) \propto Q_{{\rm n},s}$.
The center position of $i-$th nucleon, $(\bm{b}_{\perp,i},b^-_i)$, is sampled according to the Woods-Saxon distribution,
\begin{align}
f_{ \rm ws}(x^\mp,\bxp) \propto \frac{1}{1+e^{\frac{\sqrt{(x-b_{\rm imp}/2)^2+y^2+2(x^\mp-x_{\rm c})^2/\gamma^2}-R}{a}}}
\end{align}
with a nucleus radius $R$, a thinness of a nucleus surface $a$ and an impact parameter $b_{\rm imp}$.
Then, the color charge density of a nucleus with an atomic number $A$ has the following event average,
\begin{align}
&\average{ \rho^{(1/2)a}(x^\mp,\bxp)\rho^{(1/2)b}(x'^\mp,\bxp')}_{\rm eve} \nonumber\\
&= \delta^{a,b} \sum^A_{i=1} \left[g^2\mu(\frac{x^\mp+x'^\mp}{2}-b^\mp_i,\frac{\bxp+\bxp'}{2}-\bbpi)\right]^2 \nonumber\\
&\times N_{\rm 2D}(\bxp-\bxp',l_\perp) N_{\rm 1D}(x^\mp - x'^\mp,l_\mp)\ .\label{Eq:rho_tot}
\end{align}
In the limit where $R_{\rm n}$ is infinitely small and $A$ is infinitely large, Eq.~\refeq{Eq:rho_tot} becomes 
\begin{align}
&\average{ \rho^{(1/2)a}(x^\mp,\bxp)\rho^{(1/2)b}('x^\mp,\bxp') }_{\rm eve} \nonumber\\
&= \delta^{a,b} A f_{ \rm ws}(\bxp,x'^\mp) \left( g^2\bar{\mu} \right)^2 \delta(x^\mp-x'^\mp)\delta(\bxp-\bxp')\ .
\end{align}
This way of determining the color charge density for each nucleon is simple and does not take into account fluctuations of $\rho$ using the knowledge of high-energy QCD.
There are more sophisticated ways that can give more realistic determination of the color charge density. The IP-glasma model is the most famous way~\cite{IPglasma}, in which the color charge density for each nucleon is determined by the saturation scale based on the IP-sat model~\cite{IPsat1,IPsat2}. Aside from this, there is another way in which the color charge density of each nucleon is determined from the transverse momentum-dependent (TMD) gluon distribution parametrized by the GBW model~\cite{3DglasmaSS}.

We use the set of parameters listed in Table.~\ref{Tab2} chosen to describe the Au-Au collisions at $\sqrt{s}=200$ GeV. To simulate a central collision, we choose the impact parameter to be zero, $b_{\rm imp}=0$. The parameter $g^2\bar{\mu}$ is taken such that $g^2\mu_{\rm 2D, c}=10Q_{\rm n,s}$.
The infrared cutoff $m$ is introduced in the same manner as done in the previous section and is taken as $0.2$ GeV as the QCD scale.
In this setup, the energy density of the glasma generated in the central collision at $\tau=1$ fm/c and $\eta=0$ in the $x\sim y \sim 0$ region is found to be about $800$ GeV/fm$^3$. The longitudinal correlation length $l_\mp$ is taken to be the maximal value $R_{\rm n}/\sqrt{3}\gamma$ which means that the longitudinal correlation only comes from the longitudinal shape of a nucleon.
This paper focuses on studying the recoil effect of the dynamical current on the glasma, and we leave a detailed study of the effect of the longitudinal correlation on the $3+1$D glasma evolution for future works.
\begin{table}[htbp]
\caption{Parameters used in Sec.~\ref{Sec:result2}}
\label{Tab2}
\begin{center}
\begin{tabular}{r|c}
\hline\hline
$L_\perp$           & $674$                                     \\
$L_{\teta}$         & $1792$                                    \\
$a_\perp$           & $134/R$                                   \\
$a_{\teta}$         & $8/L_{\teta}$                             \\
$\gamma$            & $108$                                     \\
$A$                 & $197$                                     \\
$R$                 & $6.38$ (fm/c)                             \\
$a$                 & $0.535$ (fm/c)                            \\
$R_{\rm n}$         & $1.01$ (fm/c)                             \\
$Q_{{\rm n},s}$     & $0.5$ (GeV)                               \\
$g^2\mu_{\rm 2D,c}$ & $10 Q_{{\rm n},s}$                        \\
$l_\perp$           & $2.5/Q_{{\rm n},s}$                       \\
$l_\mp$             & $R_{\rm n}/(\sqrt{3}\gamma)$              \\
$m$                 & $0.2$ (GeV)                               \\
$b_{\rm imp}$       & $0, R$                                    \\
$\ttau_{\rm ini}$   & $0.1 a_\perp$                             \\
$x_{\rm c}$         & $\ttau_{\rm init}/\sqrt{2} + 1.8R/\gamma$ \\
\hline\hline
\end{tabular}
\end{center}
\end{table}

Note that the subtraction method used in the previous section is not used in this section. In contrast to the calculations in the previous section, the change of the CYM field of the nucleus before and after the collision of this section is not negligibly small.
As a result, the subtracted EM tensor defined in Eq.~\refeq{Eq:subt} cannot be regarded as the EM tensor consisting of only glasma contribution.

\begin{figure}[htbp]
\begin{center}
\includegraphics[width=0.7\linewidth]{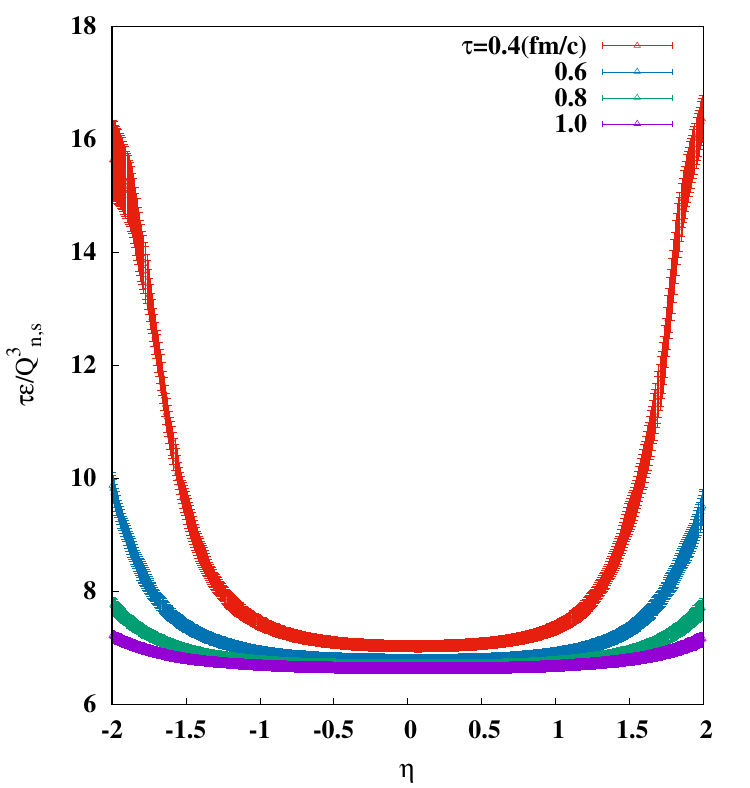}
\end{center}
\caption{The transverse-plane averaged energy density $\varepsilon$, normalized by the proper time and the saturation scale of the nucleon, at $\tau({\rm fm/c})=0.4, 0.6, 0.8$ and $1.0$.
The results are averaged over $10$ independent simulations.}
\label{Fig:ene}
\end{figure}

We show the dynamical evolution of the energy density in the central collision.
Here the energy density in the Milne coordinates is given by the energy-momentum tensor $T^{\tau\tau}$, and it is obtained from $T^{\ttau\ttau}, T^{\teta\teta}$ and $T^{\ttau\teta}$ using the general coordinates transformation,
\begin{align}
T^{\tau\tau} 
&= \frac{1}{\ttau^2-2\ttau \bar{x}_{\rm c}\cosh{\teta}+\bar{x}^2_{\rm c}}\Big[
 \left(\ttau-\bar{x}_{\rm c}\cosh{\teta}\right)^2 T^{\ttau\ttau}\nonumber\\
&+\left(\bar{x}_{\rm c}\sinh{\teta}\right)^2 \ttau^2 T^{\teta\teta}
 -\bar{x}_{\rm c}\sinh{\teta}\left(\ttau-\bar{x}_{\rm c}\cosh{\teta}\right)\ttau T^{\ttau\teta}\Big]\ .
\end{align}
In Fig.~\ref{Fig:ene}, we show the $\eta$ dependence of the energy density averaged over the transverse plane,
\begin{align}
\varepsilon \equiv \frac{\int dxdy T^{\tau\tau}}{V_\perp}\ ,
\end{align}
at different proper times in the central collisions.
The results shown in Fig.~\ref{Fig:ene} are averaged over $10$ events, and they are normalized by the proper time and the saturation scale of the nucleon.
It is found that the normalized energy density $\tau\varepsilon/Q^4_{{\rm n},s}$ rises in the large $\eta$ region.
This behavior can be interpreted as the contribution from the outgoing nuclei.
On the other hand, $\tau\varepsilon/Q^4_{{\rm n},s}$ is found to be nearly a constant function of $\eta$ in the rapidity region 
where two colliding nuclei have already passed, which indicates that the glasma created in this simulation is nearly boost-invariant. It should be noted that, compared to the calculations in the previous section, establishing the scaling law $\varepsilon \propto \tau^{-1}$ is uncertain.

\section{Summary}\label{Summary}
We have proposed a new numerical method for $3+1$D glasma simulation in Milne coordinates. 
In this method, the initial condition of the classical Yang-Mills (CYM) field and $3$D classical color current is prepared at the time before the collision of the two nuclei occurs. 
Then, the dynamical evolution of the CYM field and classical color current is tracked during the process that the two nuclei collide and pass through each other by solving the discretized evolution equations. 
Our numerical calculation is performed in the Milne coordinates ($\ttau,\teta$) where the collision has not yet occurred at $\ttau=\ttau_{\rm ini}$. Thus, the Milne coordinates we use differ from the usual Milne coordinates ($\tau,\eta$) where the center positions of two nuclei coincide at $\tau=0$. However, the physical quantities presented in the usual Milne coordinates, such as $T^{\tau\tau}$, can be obtained from that in our modified Milne coordinates by a general coordinate transformation.

Our method is a new simulation method of the $3+1$D glasma evolution with the dynamical color current. 
The difference between our method and previous methods~\cite{3DglasmaCPIC,3DglasmaCPIC2,3DglasmaCPIC3,3DglasmaSS,3DglasmaAnalytic} is that our glasma simulation is done in the modified Milne coordinates as mentioned above, while the previous simulations are done in the Minkowski coordinates. Since the numerical simulations on a finite lattice in Milne coordinates correspond to a longitudinally expanding system in terms of the Minkowski coordinates because $z=\tau \sinh{\eta}$, numerical simulations with our Milne coordinates need much less numerical resources than the simulations in Minkowski coordinates. 
Reduction of the numerical resource is important in the actual application since tracking the dynamical evolution of the $3$D glasma requires huge numerical resources.

In Sec.~\ref{Sec:result1}, we first confirmed that two relations derived from the continuity equation of the EM tensor are not violated in the actual simulations, which indicates that the discretization effect on the dynamics is tiny and is well under control. Then, we checked the consistency of our results and the results shown in Ref.~\cite{3DglasmaSS}, using the same setup as Ref.~\cite{3DglasmaSS}. As a result, the transverse pressure $P_\perp$ calculated in our method completely agrees with their result. The most important thing to note is that we can reproduce the their result using about $4.5$ smaller number of grids in the longitudinal direction. The number of grids in the $z$ direction in their calculation is $2048$, while the number of grids in the $\teta$ direction in our calculation is $448$.
In addition, we have explicitly shown that the transverse pressure decrease as $P_\perp \propto \tau^{-1}$, which is treated as the assumption in Ref.~\cite{3DglasmaSS}. On the other hand, the energy density in the local rest frame $\varepsilon_{{\rm LRF}}$ calculated in our method do not fully agree with the results shown in Ref.~\cite{3DglasmaSS}.
Our calculation has shown that it is about $2$ times the transverse pressure, which means that the transverse pressure $P_\perp$ is much larger than the longitudinal pressure in the local rest frame, $P_{{\rm LRF},L}=\varepsilon_{{\rm LRF}}-2P_\perp$.  

In Sec.~\ref{Sec:result2}, we show the numerical results using the initial conditions that describe the central Au-Au collisions at $\sqrt{s}=200$ GeV. 
The energy density of the glasma is found to be almost boost-invariant near the midrapidity region. It should be noted that, compared to the calculations with the setup consistent with the Ref.~\cite{3DglasmaSS}, establishing the scaling behavior, $\varepsilon \propto \tau^{-1}$, is uncertain.

As explained above, we have shown that our method costs much lower numerical resources, and thus it can be used for practical event-by-event simulations within reasonable simulation time. The natural future directions are then to investigate other observables such as angular momentum, topological charges, to study the thermalization of the glasma, and to provide the initial conditions for hydrodynamic simulations for the comparison of the obtained results with experimental results. Aside from these, another possible direction is to make the initial condition more realistic using the phenomenological model based on the high-energy QCD, such as IP-sat model~\cite{IPsat1,IPsat2}.

\section*{Acknowledgments}
The authors thank M. Leuthner, S. Schlichting, and P. Singh for useful discussions. The authors also acknowledge Yukawa-21 for providing the computational resources. This work is supported by the National Natural Science Foundation of China (Grant No. 12147101, No. 12225502, and No. 12075061), the National Key Research and Development Program of China (Grant No. 2022YFA1604900), and the Natural Science Foundation of Shanghai (Grant No. 20ZR1404100).

\appendix

\section{Wilson line on a lattice}\label{App:Wilson}
To obtain the initial condition of the CYM field and the classical color current on the lattice, shown in Sec.~\ref{Sec:3discr}, we have to evaluate the Wilson line at ($\ttau_{\rm ini},\teta$) numerically,
\begin{align}
V^{(1/2)\dagger}_x
= P_{x^\mp} \exp{\left[ -i \int^{x^\mp}_{-\infty} dx'^\mp \partial^{-2}_\perp 
\rho^{(1/2)}_{\rm cov}(x'^\mp,\bxp) \right]}\ .\label{Eq:Wilson2}
\end{align}
In our setup, the color charge density $\rho^{(1/2)}_{\rm cov}$ doesn't exist at the sufficiently small $x^\mp$, and thus we can replace the lower bound of the integral in Eq.~\refeq{Eq:Wilson2} with a small value $x_{\rm low}$ such that $\rho^{(1/2)}|_{x^\mp=x_{\rm low}} \sim 0$.
Then, for the numerical evaluation of Eq.~\refeq{Eq:Wilson2}, we divide the exponential function in Eq.\refeq{Eq:Wilson2} into many small parts,
\begin{align}
V^{(1/2)\dagger}_x 
=&W^{(1/2)}_x\Big|_{x^{\teta}=\teta\pm \frac{1}{2}\Delta \teta} W^{(1/2)}_x\Big|_{x^{\teta}=\teta\pm \frac{3}{2}\Delta \eta} \nonumber\\
 &\quad\cdots\ W^{(1/2)}_x\Big|_{x^{\teta}=\mp \ln{\frac{\sqrt{2}x_{\rm low}}{\ttau_{\rm ini}}} },\label{Eq:VW}
\end{align}
where $W^{(1/2)}$ is spaced with interval $\Delta \teta$ in $\teta$ coordinates and is given in the following expression,
\begin{align}
&W^{(1/2)}_x\nonumber\\&=\exp \Bigl[ 
-i 
\left| x^\mp|_{x^{\teta}=\teta + \frac{1}{2}\Delta \teta} - x^\mp|_{x^{\teta}=\teta - \frac{1}{2}\Delta \teta} \right|
\partial^{-2}_\perp \rho^{(1/2)}_{\rm cov}|_{x^{\teta}=\teta}
\Bigr]\ .
\end{align}
Here $\partial^{-2}_\perp \rho^{(1/2)}_{\rm cov}$ is obtained by the discrete Fourier transform as shown in Eq.~\refeq{Eq:rho_fourier}.
The interval $\Delta \teta$ should be small enough to converge the evaluated Wilson line.

\section{Discretization of time direction}\label{App:evo}
We show here how to solve the evolution equations shown in Sec.~\ref{Sec:3discr} by the difference method.
In the actual calculations, we use the time variable $\theta=\ln{\ttau}$ instead of $\ttau$.
Because of the relation $\partial_\theta = \ttau \partial_{\ttau}$, we can solve the evolution equations efficiently in the small $\tau$ region where the numerical calculation is more severe.
The discretized classical equation of motion with the step size $a_\theta$ is given by,
\begin{align}
&U_{i,x}|_{\theta=\theta_{\rm ini} + (n+2)a_\theta}
= e^{\frac{gi a_\theta g_{ii} E^i}{a_{\teta}}}|_{\theta=\theta_{\rm ini} + (n+1)a_\theta}    U_{i,x}|_{\theta=\theta_{\rm ini} + n a_\theta}\ ,\\
&E^i_x|_{\theta=\theta_{\rm ini} + (n+2)a_\theta}= E^i_x|_{\theta=\theta_{\rm ini} + n a_\theta}\nonumber\\
&- a_\theta a_{\teta} \Bigl\{ \frac{i \ttau}{2g } \sum_i g^{ii} g^{jj} \left[W_{ij,x} -  U^\dagger_{j,x-\hat{j}}W_{ij,x-\hat{j}}U_{j,x-\hat{j}} \right]\nonumber\\
&+ \delta^{i \teta} \left[ \tilde{J}^{(1)}_{\rm L,x} - \tilde{J}^{(2)}_{\rm L,x} \right]\Bigr\} \Big|_{\theta=\theta_{\rm ini} + (n+1)a_\theta}\ ,
\end{align}
where $\theta_{\rm ini}=\ln{\ttau_{\rm ini}}$ is $\theta$ at the initial proper time and $n$ is the time step.
The continuity equations are discretized as,
\begin{align}
&\tilde{J}^{(1/2)}_x|_{\theta=\theta_{\rm ini} + (n+2)a_\theta}           = \tilde{J}^{(1/2)}_x|_{\tau=n a_\theta} \nonumber\\
&\ \ \ \ \ \mp\frac{a_\theta}{a_{\tilde{\eta}}} \left[ \tilde{J}^{(1/2)}_{{\rm L},x} - U^\dagger_{\tilde{\eta},x-\hat{\teta}} \tilde{J}^{(1/2)}_{{\rm L},x-\hat{\teta}} U_{\teta,x-\hat{\teta}} \right]|_{\theta=\theta_{\rm ini} + (n+1)a_\theta}\ ,\\
&\tilde{J}^{(1/2)}_{{\rm L},x}|_{\theta=\theta_{\rm ini} + (n+2)a_\theta} = \tilde{J}^{(1/2)}_{{\rm L},x}|_{\theta=\theta_{\rm ini} + n a_\theta} \nonumber\\
&\ \ \ \ \ \mp\frac{a_\theta}{a_{\teta}}        \left[ U_{\teta,x} \tilde{J}^{(1/2)}_{x+\hat{\teta}} U^\dagger_{\teta,x} - \tilde{J}^{(1/2)}_x \right]|_{\theta=\theta_{\rm ini} + (n+1)a_\theta}\ .
\end{align}
In the actual calculations, we solve these discretized evolution equations by the leap-flog method, which is convenient for describing the Hamilton dynamics.

\section{Longitudinal pressure in setup for Sec.~\ref{Sec:result1}}\label{App:lon}
\begin{figure}[htbp]
\begin{center}
\includegraphics[width=0.7\linewidth]{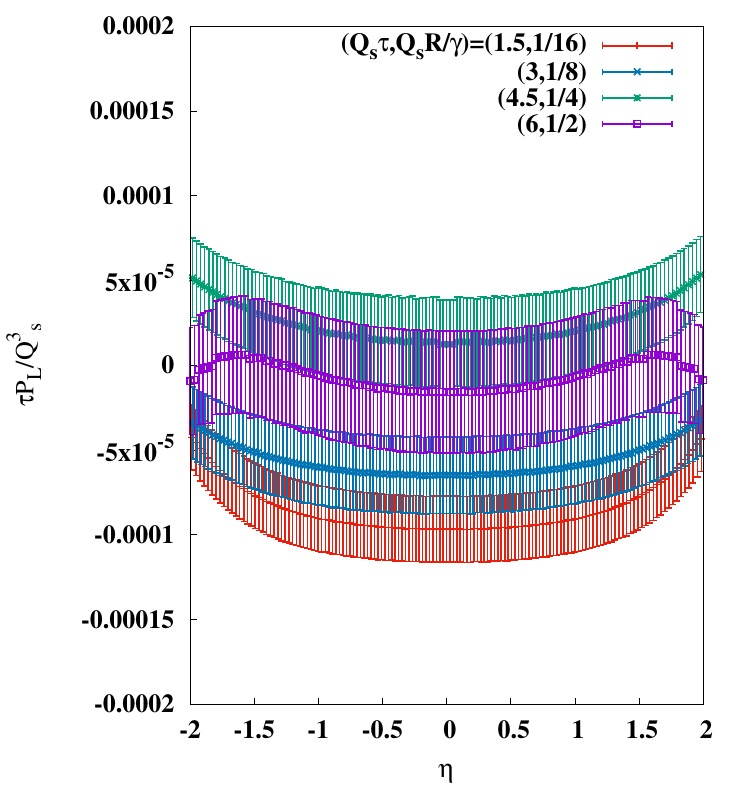}
\end{center}
\caption{
The $\eta$ dependence of the longitudinal pressure normalized by the proper time and the saturation scale, $\tau P_{\rm L}/Q^3_s$.
All results shown here are calculated from the same simulations as shown in Fig.~\ref{Fig:pt} and Fig.~\ref{Fig:e}.
}
\label{Fig:pl}
\end{figure}

In this appendix, we show the longitudinal pressure is negligibly small, compared to the transverse pressure shown in Fig.~{Fig:pt}, 
which means that the whole system expands like a rarefied gas in the longitudinal direction.
The longitudinal pressure is defined as 
\begin{align}
P_{\rm L} \equiv \frac{ \int d^2x_\perp \tau^2 T^{\eta\eta}_{\rm sub} }{2V_\perp}\ .
\end{align}
Here the energy momentum tensor $T^{\eta\eta}_{\rm sub}$ is calculated via the general coordinates transformation,
\begin{align}
&\tau^2 T^{\eta\eta}_{\rm sub}
= \frac{1}{\ttau^2-2\ttau \bar{x}_{\rm c}\cosh{\teta}+\bar{x}^2_{\rm c}}\Big[
 \left(\bar{x}_{\rm c}\sinh{\teta}\right)^2 T^{\ttau\ttau}_{\rm sub} \nonumber\\
&+\left(\ttau-\bar{x}_{\rm c}\cosh{\teta}\right)^2 \ttau^2 T^{\teta\teta}_{\rm sub}
 -\bar{x}_{\rm c}\sinh{\teta}\left(\ttau-\bar{x}_{\rm c}\cosh{\teta}\right)\ttau T^{\ttau\teta}_{\rm sub} \Big]\ .
\end{align}
Figure~\ref{Fig:pl} shows the $\eta$ dependence of $P_{\rm L}$ normalized by the proper time and the saturation scale, $\tau P_{\rm L}/Q^3_s$, for different thicknesses, $Q_sR/\gamma=1/2, 1/4, 1/8$ and $1/16$.
Comparing Fig.~\ref{Fig:pt} and Fig.~\ref{Fig:pl}, it is found that the longitudinal pressure is much smaller than the transverse pressure in the wide rapidity range, $-2<\eta<2$.

\section{Scaling bahavior}\label{App:scaling}
In this appendix, we discuss the scaling behavior obtained in Fig.~\ref{Fig:pt}, $P_\perp \propto \tau^{-1}$.
First, let us see the continuity equation for $T^{\tau\tau}$ without the dynamical current, 
\begin{align}
\frac{1}{\tau}\left\{ \partial_\tau \left[\tau T^{\tau\tau}\right] + \tau^2 T^{\eta\eta}\right\}+\partial_1 T^{1\tau}+\partial_2 T^{2\tau}+\partial_\eta T^{\eta\tau}
=0\ .
\end{align}
The current term is neglected here since the nuclei have already passed through in the time region shown in Fig.~\ref{Fig:pt} and Fig.~\ref{Fig:e}.
By the integration over the transverse plane, the continuity equation leads to the evolution equation of the sum of the transverse and longitudinal pressure,
\begin{align}
\partial_\tau \left[\tau (P_\perp+P_{\rm L})\right]=-P_{\rm L}-\tau \partial_\eta \tau^{\eta\tau}\ .\label{Eq:cont_wo_current}
\end{align}
where $\tau^{\eta\tau} \equiv \int dxdy T^{\eta\tau}/V_\perp$, the longitudinal pressure is defined as $P_{\rm L} = \int dxdy\tau^2  T^{\eta\eta} / V_\perp$, 
To obtain Eq.~\refeq{Eq:cont_wo_current}, we use the relation $T^{\tau\tau} = T^{11}+T^{22}+T^{\eta\eta}$ resulting from the conformal symmetry of the CYM theory.
Since the longitudinal pressure is quite smaller than the transverse pressure, as shown in appendix~\ref{App:lon}, 
we can drop $P_{\rm L}$ and obtain the evolution equation of the transverse pressure as
\begin{align}
\partial_\tau \left[\tau P_\perp \right]=-\tau \partial_\eta \tau^{\eta\tau}\ .\label{Eq:evo_Pperp}
\end{align}
Therefore the realization of the scaling behavior indicates that the derivative term $\tau \partial_\eta \tau^{\eta\tau}$ in Eq.~\refeq{Eq:evo_Pperp} is negligible as well as $P_{\rm L}$.

\section{Derivation of $2$-point correlation function of color charge density of a single nucleon}\label{App:rho}
In this appendix, we show the detailed calculation process to obtain Eq.~\refeq{Eq:rho_i_v2}.
First, let us substitute Eq.~\refeq{Eq:gamma_i} into Eq.~\refeq{Eq:rho_i},
\begin{align}
&\average{ \rho^{(1/2)a}_i(x^\mp,\bxp)\rho^{(1/2)b}_i(x'^\mp,\bxp') }_{\rm eve} \nonumber\\
&=
\delta^{a,b} 
\left( g^2\bar{\mu} \right)^2
2\pi\left(\frac{2R^2_{\rm n}}{3}+\sigma^2_\mp\right)
\sqrt{ 2\pi\left(\frac{R^2_{\rm n}}{3\gamma^2}+\sigma^2_\perp\right) } \nonumber\\
&\times N_{\rm 1D}(x^\mp -b^\mp_i,\frac{R_{\rm n}}{\sqrt{6}\gamma})N_{\rm 2D}(\bxp -\bbpi,\frac{R_{\rm n}}{\sqrt{3}})\nonumber\\
&\times N_{\rm 1D}(x'^\mp-b^\mp_i,\frac{R_{\rm n}}{\sqrt{6}\gamma})N_{\rm 2D}(\bxp'-\bbpi,\frac{R_{\rm n}}{\sqrt{3}})\nonumber\\
&\times N_{\rm 1D}(x^\mp - x'^\mp,\sigma_\mp)N_{\rm 2D}(\bxp-\bxp',\sigma_\perp)\ .\label{Eq:rho_i_v25}
\end{align}
The product of Gussian functions of $x^\mp$ and $x'^\mp$ in the right-hand of Eq.~\refeq{Eq:rho_i_v25} can be transformed into the product of Gussian functions of $x^\mp+x'^\mp$ and $x^\mp-x'^\mp$ as
\begin{align}
&\average{ \rho^{(1/2)a}_i(x^\mp,\bxp)\rho^{(1/2)b}_i(x'^\mp,\bxp') }_{\rm eve} \nonumber\\
&=
\delta^{a,b} 
\left( g^2\bar{\mu} \right)^2
2\pi\left(\frac{2R^2_{\rm n}}{3}+\sigma^2_\perp\right)
\sqrt{ 2\pi\left(\frac{R^2_{\rm n}}{3\gamma^2}+\sigma^2_\mp\right) } \nonumber\\
&\times N_{\rm 1D}(\frac{x^\mp+x'^\mp}{2}-b^\mp_i,\frac{R_{\rm n}}{2\sqrt{3}\gamma}) N_{\rm 1D}(x^\mp - x'^\mp,\frac{R_{\rm n}}{\sqrt{3}\gamma})\nonumber\\
&\times N_{\rm 2D}(\frac{\bxp+\bxp'}{2}-\bbpi,\frac{R_{\rm n}}{\sqrt{6}}) N_{\rm 2D}(\bxp - \bxp',\sqrt{\frac{2}{3}}R_{\rm n})\nonumber\\
&\times N_{\rm 1D}(x^\mp - x'^\mp,\sigma_\mp)N_{\rm 2D}(\bxp-\bxp',\sigma_\perp)\ .\label{Eq:rho_i_v3}
\end{align}
Here we use the following relations,
\begin{align}
&N_{\rm 1D}(x^\mp -b^\mp_i,\frac{R_{\rm n}}{\sqrt{6}\gamma})
 N_{\rm 1D}(x'^\mp-b^\mp_i,\frac{R_{\rm n}}{\sqrt{6}\gamma})\nonumber\\
&=
N_{\rm 1D}(\frac{x^\mp+x'^\mp}{2}-b^\mp_i,\frac{R_{\rm n}}{2\sqrt{3}\gamma}) N_{\rm 1D}(x^\mp - x'^\mp,\frac{R_{\rm n}}{\sqrt{3}\gamma})\nonumber\\
&N_{\rm 2D}(\bxp -\bbpi,\frac{R_{\rm n}}{\sqrt{3}}) N_{\rm 2D}(\bxp'-\bbpi,\frac{R_{\rm n}}{\sqrt{3}})\nonumber\\
&=
N_{\rm 2D}(\frac{\bxp+\bxp'}{2}-\bbpi,\frac{R_{\rm n}}{\sqrt{6}}) N_{\rm 2D}(\bxp - \bxp',\sqrt{\frac{2}{3}}R_{\rm n})\ .
\end{align}
A number of Gaussian functions in the right-hand of Eq.~\refeq{Eq:rho_i_v3} can be reduced by using the relation,
\begin{align}
& N_{\rm 1D}(x^\mp - x'^\mp,\frac{R_{\rm n}}{\sqrt{3}\gamma})
  N_{\rm 1D}(x^\mp - x'^\mp,\sigma_\mp)\nonumber\\
&=\frac{1}{\sqrt{2\pi(R^2_{\rm n}/(3\gamma^2)+\sigma^2_\mp)}} N_{\rm 1D}(x^\mp - x'^\mp,l_\mp)\nonumber\\
& N_{\rm 2D}(x_\perp - x'_\perp,\sqrt{\frac{2}{3}}R_{\rm n})
  N_{\rm 2D}(x_\perp - x'_\perp,\sigma_\perp)\nonumber\\
&=\frac{1}{2\pi(2R^2_{\rm n}/3+\sigma^2_\perp)} N_{\rm 2D}(x_\perp - x'_\perp,l_\perp)\ ,
\end{align}
where the correlation lengths $l_\mp$ and $l_\perp$ are defined in Eq.~\refeq{Eq:lmp} and Eq.~\refeq{Eq:lp}, respectively.
Then, Eq.~\refeq{Eq:rho_i_v3} reads Eq.~\refeq{Eq:rho_i_v2},
\begin{align}
&\average{ \rho^{(1/2)a}_i(x^\mp,\bxp)\rho^{(1/2)b}_i(x'^\mp,\bxp') }_{\rm eve} \nonumber\\
&= \delta^{a,b} \left( g^2\bar{\mu} \right)^2 
        N_{\rm 1D}(\frac{x^\mp+x'^\mp}{2}-b^\mp_i,\frac{R_{\rm n}}{\sqrt{3}\gamma}) N_{\rm 1D}(x^\mp - x'^\mp,l_\mp)\nonumber\\
&\times N_{\rm 2D}(\frac{\bxp+\bxp'}{2}-\bbpi,\sqrt{\frac{2}{3}}R_{\rm n})          N_{\rm 2D}(\bxp-\bxp',l_\perp)\ .
\end{align}

\end{document}